\documentclass[12pt]{article}
\usepackage{latexsym}
\usepackage{amsmath,,calrsfs}
\usepackage{amsfonts}
\usepackage{amssymb}
\usepackage{amscd}
\usepackage{bbm}
\usepackage{fancybox}
\usepackage{cite}
\usepackage{amsmath,amsfonts,amsbsy}
\usepackage{pstricks,pst-node}
\usepackage[small,bf,hang]{caption2}
\usepackage{graphicx}
\usepackage{epsfig}
\usepackage{psfrag}
\usepackage{comment}

\usepackage{float}

\psset{unit=1.3cm,linewidth=.5pt,radius=.2}  

\usepackage{multirow}                     
\usepackage{float}                          
\usepackage{lscape}                         
\usepackage{bm}

\newcommand{\beq}{\begin{equation}}
\newcommand{\eeq}{\end{equation}}
\newcommand{\bi}{\begin{itemize}}
\newcommand{\ei}{\end{itemize}}
\newcommand{\bt}{\begin{tabular}}
\newcommand{\et}{\end{tabular}}
\newcommand{\bc}{\begin{center}}
\newcommand{\ec}{\end{center}}

\newcommand{\dslash}{\partial \hskip -.24truecm  / }

\newcommand{\ket}[1]{|#1\rangle}

\newcommand{\ft}[2]{{\textstyle {\frac{#1}{#2}} }}

\newcommand{\be}{\begin{equation}}
\newcommand{\ee}{\end{equation}}
\newcommand{\bea}{\begin{eqnarray}}
\newcommand{\eea}{\end{eqnarray}}
\newcommand{\ba}{\begin{array}}
\newcommand{\ea}{\end{array}}

\def\bbox{{\,\lower0.9pt\vbox{\hrule \hbox{\vrule height 0.2 cm
\hskip 0.2 cm \vrule height 0.2 cm}\hrule}\,}}
\newcommand{\dsl}{\pa \kern-0.5em /}

\def\eq#1{(\ref{#1})}

\makeatletter \@addtoreset{equation}{section} \makeatother

\def\slashchar#1{\setbox0=\hbox{$#1$}           
   \dimen0=\wd0                                 
   \setbox1=\hbox{/} \dimen1=\wd1               
   \ifdim\dimen0>\dimen1                        
      \rlap{\hbox to \dimen0{\hfil/\hfil}}      
      #1                                        
   \else                                        
      \rlap{\hbox to \dimen1{\hfil$#1$\hfil}}   
      /                                         
   \fi}

\def\eq#1{(\ref{#1})}

\begin{document}

\begin{titlepage}
\begin{center}

\hfill UG-10-75 \\ \hfill MIT-CTP-4163\\ \hfill DAMTP-2010-53\\

\vskip 1.5cm

{\Large \bf  On Maximal Massive 3D Supergravity}

\vskip 1cm

{\bf Eric A.~Bergshoeff\,$^1$, Olaf Hohm\,$^2$, Jan Rosseel\,$^1$\\[0.5ex] and Paul K.~Townsend\,$^3$} \\

\vskip 25pt

{\em $^1$ \hskip -.1truecm Centre for Theoretical Physics,
University of Groningen, \\ Nijenborgh 4, 9747 AG Groningen, The
Netherlands \vskip 5pt }

{email: {\tt E.A.Bergshoeff@rug.nl, j.rosseel@rug.nl}} \\

\vskip 15pt

{\em $^2$ \hskip -.1truecm Center for Theoretical Physics,
Massachusetts Institute of Technology, \\ Cambridge, MA 02139, USA
\vskip 5pt }

{email: {\tt ohohm@mit.edu}} \\

\vskip 15pt

{\em $^3$ \hskip -.1truecm Department of Applied Mathematics and Theoretical Physics,\\ Centre for Mathematical Sciences, University of Cambridge,\\
Wilberforce Road, Cambridge, CB3 0WA, U.K. \vskip 5pt }

{email: {\tt P.K.Townsend@damtp.cam.ac.uk}} \\

\end{center}

\vskip 0.5cm

\begin{center} {\bf ABSTRACT}\\[3ex]
\end{center}

We construct, at the linearized level,  the  three-dimensional (3$D$) $\mathcal{N}=4$ supersymmetric ``general massive supergravity''  and the
maximally supersymmetric $\mathcal{N}=8$ ``new massive supergravity''.  We also construct the maximally supersymmetric linearized $\mathcal{N}=7$ topologically massive supergravity, although we expect $\mathcal{N}=6$ to be maximal at the non-linear level.








\end{titlepage}

\newpage
\setcounter{page}{1} \tableofcontents

\newpage


\section{Introduction}

The ``on-shell''  $\mathcal{N}$-extended one-particle
supermultiplets available for massless particles of four-dimensional
(4D) field theories are well-known. Generically, these
supermultiplets will appear in CPT-dual pairs but  there are some
special supermultiplets that are CPT self-dual; these are unique for
a given choice of maximal spin, and have  the property that
$\mathcal{N}$ is maximal for that spin. The corresponding field
theories generally have improved ultra-violet (UV) behaviour. For
example, $\mathcal{N}=4$ is maximal for maximum spin 1, and there is
a unique $\mathcal{N}=4$ self-dual supermultiplet, which is realized
by  the UV-finite $\mathcal{N}=4$ super-Yang-Mills (SYM) theory.
Similarly, $\mathcal{N}=8$ is maximal for maximum spin 2, and there
is a unique $\mathcal{N}=8$ self-dual supermultiplet, which is
realized by $\mathcal{N}=8$ supergravity; the UV status of this
non-renormalizable theory  is still a matter of dispute but it is
certainly ``improved''; see \cite{Dixon:2010gz} for a recent review.

This paper is motivated by the observation that there is a
one-to-one correspondence between the massless on-shell
supermultiplets of $\mathcal{N}$-extended 4D supersymmetry and the
{\it massive}  on-shell  supermultiplets of $\mathcal{N}$-extended
three-dimensional (3D) supersymmetry.
The $\mathcal{N}$ 3D Majorana spinor supercharges $Q^i$, $i=1,\cdots,\mathcal{N}$, obey the anticommutation relations
 \bea\label{superPoincare}
  \{ Q_{\alpha}^{i},Q_{\beta}^{j}\} \ = \ 2(\gamma^{\mu}C)_{\alpha\beta}\,
  P_{\mu}\,\delta^{ij}\;,
 \eea
where $P_\mu$ is the three-momentum, $\gamma^\mu$ are the 3D Dirac
matrices and $C$ is the charge conjugation matrix. Choosing a real
representation of the Dirac matrices with $C = \gamma^0$ and
choosing the rest frame three-momentum $P_\mu = (-M,0,0)$ for a
particle of mass $M$, one finds that the following combinations
 \bea
  (a^{i})^{\dagger} \ = \ \ft12\left(Q_{1}^{i}+iQ_{2}^{i}\right)\;,
  \qquad
  a^{i} \ = \ \ft12\left(Q_1^{i}-iQ_{2}^{i}\right)\;,
 \eea
obey the anticommutation relations
 \bea\label{massiveosc}
  \{ a^{i},(a^j)^{\dagger}\} \ = \
  M\delta^{ij}\;, \qquad
  \{ a^{i},a^j \} \ = \ \{ (a^{i})^{\dagger},(a^{j})^{\dagger}\} \ =
  \ 0\;.
 \eea
The operators $a$ and $a^{\dagger}$ can thus be seen as raising and
lowering operators. Moreover, it can be checked that they increase
or decrease the space-time helicity by 1/2 (see
e.g.~\cite{Hohm:2004rc}). The construction of the supermultiplets is
thus straightforward. Starting from a `Clifford vacuum'
$\ket{\Omega}_{j}$ with helicity $j$ one can act ${\cal N}$ times
with the raising operators $a^{\dagger}$. This leads to a massive
multiplet with $2^{\cal N}$ states of helicities ranging from $j$ to
$j+{\cal N}/2$, as shown in Table \ref{multiplets} for $j+{\cal
N}/2=2$. Only the multiplets containing the $+2$ helicities are
shown for $\mathcal{N}=1,\cdots, 7$. Parity flips the helicities and
hence takes these into multiplets containing a state of helicity
$-2$ but not $+2$. The $\mathcal{N}=8$ multiplet is exceptional
because  it contains both helicities $+2$ and $-2$ and is therefore
``parity self-dual''.  One sees that this is formally the same
construction as massless particle supermultiplets of
four-dimensional $\mathcal{N}$-extended supersymmetry\footnote{If
the anticommutator (\ref{superPoincare})  is modified to allow
additional (non-central) charges then one can find additional,
parity-preserving,  multiplets that  do not  correspond to 4D
multiplets because they are acted upon by twice as many supercharges
as those discussed here \cite{Bergshoeff:2008ta}.}.  In the 4D case,
however, one must include the CPT-conjugate multiplets.

\begin{table}[t]\label{multiplets}
\caption{{\footnotesize{Some 3D massive supermultiplets}}}
\vspace{0.1cm}
\centering
\begin{tabular}{cccccccccc}
\hline\hline\\[-.2truecm]
helicity&+2&+3/2&+1&+1/2&0&--1/2&--1&--3/2&--2\\[.2truecm]
\hline\\[-.2truecm]
${\cal N}=1$&1&1&&&&&&&\\[.2truecm]
${\cal N}=2$&1&2&1&&&&&&\\[.2truecm]
${\cal N}=3$&1&3&3&1&&&&&\\[.2truecm]
${\cal N}=4$&1&4&6&4&1&&&&\\[.2truecm]
${\cal N}=5$&1&5&10&10&5&1&&&\\[.2truecm]
${\cal N}=6$&1&6&15&20&15&6&1&&\\[.2truecm]
${\cal N}=7$&1&7&21&35&35&21&7&1&\\[.2truecm]
${\cal N}=8$&1&8&28&56&70&56&28&8&1\\[.2truecm]\hline
\end{tabular}
\end{table}

This 4D/3D correspondence holds for each value of $\mathcal{N}$
although it should be appreciated that the 3D theory has half the
total number of supersymmetries  (because the minimal spinor in 3D
has half as many independent components as the minimal  spinor in
4D).  It should also be appreciated that  the one-to-one
correspondence is for supermultiplets, and not free field
theories\footnote{Interactions must be considered on a case-by-case
basis.};  this is because a CPT-dual pair in 4D corresponds to a
pair of  3D massive supermultiplets paired by parity, but locality
in 3D does not  require this pairing. This means that there can be
parity-violating massive 3D field theories that have no 4D analog;
the simplest  ($\mathcal{N}=0$) examples are ``topologically massive
electrodynamics'' (TME)
\cite{Siegel:1979fr,Schonfeld:1980kb,Deser:1981wh} and
``topologically massive gravity'' (TMG)
\cite{Deser:1981wh,Deser:1982vy}.

Thus we can expect to find (at least at the linearized level)
parity-preserving massive 3D SYM and massive 3D supergravity
theories for each of the corresponding massless 4D SYM and
supergravity theories. This expectation is realized in the SYM case
by the $\mathcal{N}$-extended 3D SYM-Higgs theories, expanded about
a Higgs vacuum in which all particles are massive. This yields  3D
theories of massive spin-$1$ particles with $\mathcal{N}=1,2,4$
supersymmetry. Parity may then be broken by the addition of a
supersymmetric Chern-Simons (CS) term, but such a term exists only
for $\mathcal{N}=1,2,3$, so $\mathcal{N}=3$ is maximal  for spin 1
if parity is violated \cite{Kao:1992ig}. This  feature can be
understood directly from the  supermultiplet structure: the
$\mathcal{N}=4$ spin-$1$ supermultiplet  is equivalent to a
parity-dual pair of $\mathcal{N}=3$ spin-$1$ supermultiplets, so any
parity violation that  is visible in the linearized theory will
split the degeneracy of this pair, thereby breaking $\mathcal{N}=4$
to $\mathcal{N}=3$.

In the supergravity case, the 4D/3D analogy leads us to expect
parity-preserving  3D supergravity theories propagating massive
graviton supermultiplets with $\mathcal{N}=1,2,3,4,5,6,8$.  Until
recently, it was far from obvious how such theories could be
constructed, but the $\mathcal{N}=0$ example of ``New Massive
Gravity'' (NMG) \cite{Bergshoeff:2009hq} has shown the way, because
it propagates precisely massive helicity $\pm 2$ modes. Furthermore,
one can add a  Lorentz Chern-Simons (LCS) term to the NMG action to
obtain a parity-violating   ``General Massive Gravity'' (GMG)  that
has both NMG and TMG as limiting cases \cite{Bergshoeff:2009hq}. The
$\mathcal{N}=1$ supersymmetric extension of these 3D gravity
theories has now been constructed
\cite{Andringa:2009yc,Bergshoeff:2010mf} as has the linearized
$\mathcal{N}=2$ extension, which nicely combines NMG with the Proca
action for spin 1, and combines both CS and LCS terms in a single
$\mathcal{N}=2$ superinvariant \cite{Andringa:2009yc}.  The main
purpose of this paper is to present results for the $\mathcal{N}>2$
massive supergravity theories.\footnote{A ``disjoint''  family of
massive 3D $\mathcal{N}=1$ supergravity theories was recently found
in  \cite{Lu:2010ct} by  dimensional reduction of a 6D theory with
curvature-squared terms. These may also have $\mathcal{N}>1$
extensions but we shall have nothing to say about that here.}

We shall work exclusively at the linearized level, leaving the
problem of interactions to future work. Even so, there are a number
of issues that we are able to address, and resolve. Firstly,  we
recall that NMG consists of the Einstein-Hilbert action, with the
``wrong'' sign, and a curvature-squared invariant constructed from
the scalar $K= G^{\mu\nu}S_{\mu\nu}$, where $G_{\mu\nu}$ is the
Einstein tensor and $S_{\mu\nu}$ the Schouten tensor.  We therefore
need an $\mathcal{N}=8$ extension of the Einstein-Hilbert action
(the lower $\mathcal{N}$  cases  can then be obtained by
truncation). This involves  coupling the $\mathcal{N}=8$ off-shell
Weyl supermultiplet \cite{Howe:1995zm} (which contains the graviton
field subject to  linearized  diffeomorphisms {\it and} linearized
Weyl rescalings)  to 8 compensating  $\mathcal{N}=8$ scalar
supermultiplets (containing only scalar and spinor fields).   Each
off-shell $\mathcal{N}=8$ scalar supermultiplet  has an infinite
number of auxiliary fields (arising from the expansion of a field
defined on  $\mathcal{N}=8$ harmonic superspace \cite{Howe:1998jw}).
These should be eliminated only  {\it after} the addition of an
$\mathcal{N}=8$ extension of the NMG K-invariant, but the
compensating supermultiplets decouple from the K-superinvariant {\it
at the linearized level} as a consequence of an  ``accidental''
linearized superconformal invariance; this allows us to trivially
eliminate all auxiliary fields of each compensating supermultiplet and
work with the simpler on-shell-supersymmetric\footnote{By
``on-shell-supersymmetric'' we mean that the equations of motion are
needed for closure of the supersymmetry algebra, due to the implicit
elimination of auxiliary fields.} $\mathcal{N}=8$ scalar
supermultiplet.

Starting with the $\mathcal{N}=8$ super-NMG model, we might expect
to be able to construct a parity-violating $\mathcal{N}=8$ super-GMG
model by the addition of  an  $\mathcal{N}=8$  super-LCS term. Let
us recall that the LCS term is, by itself, the action of 3D
conformal gravity \cite{vanNieuwenhuizen:1985cx} and that there
exists an  $\mathcal{N}=8$ superconformal gravity
\cite{Gran:2008qx}. However, the $\mathcal{N}=8$ spin-$2$
supermultiplet is equivalent to a  degenerate parity-dual pair of
$\mathcal{N}=7$ spin-$2$ supermultiplets, and this degeneracy is
lifted if parity is violated. It follows that, at best,
$\mathcal{N}=7$ is maximal for super-GMG and also for its TMG limit.
In fact, for reasons that we will give later, we believe that
$\mathcal{N}=6$ is actually maximal for the non-linear TMG and
generic GMG. However, we verify that there is  an $\mathcal{N}=7$
{\it linearized}  super-GMG theory by constructing the linearized
off-shell $\mathcal{N}=7$ super-LCS invariant; this can be added to
the  $\mathcal{N}=8$ super-NMG action because it does not couple to
the compensating supermultiplet.  We also explain why the
$\mathcal{N}=8$ super-LCS invariant cannot be similarly used to
construct an $\mathcal{N}=8$ super-GMG theory; in brief, it is
because there is no {\it off-shell-supersymmetric}  $\mathcal{N}=8$
super-LCS term.

The method of construction of the linearized $\mathcal{N}=7$, $8$
massive gravities that we have just sketched, can be applied for any
$\mathcal{N}$. As a preliminary,  we begin with a sketch of how the
construction works for each value of $\mathcal{N}$.  This involves a
determination of the off-shell Weyl supermultiplet and the
compensating multiplets needed for the construction of  the
linearized $\mathcal{N}$-extension of the Einstein-Hilbert term. For
$\mathcal{N}\le3$, not only does a single scalar supermultiplet
suffice but there is also an $SO(\mathcal{N})$ singlet in this
multiplet than can be identified with the compensating scalar for
the local scale invariance of the Weyl multiplet. The situation for
$\mathcal{N}=4$ is quite different. In this case, the scalar
multiplet is the 3D version of the 4D hypermultiplet, but there are
two distinct versions of it in 3D, and both are needed. Furthermore,
the local scale compensator is necessarily composite because neither
hypermultiplet contains a singlet of the R-symmetry group. In
several respects, the $\mathcal{N}=4$ case is similar to the
$\mathcal{N}=8$ case, but much simpler, so we discuss
$\mathcal{N}=4$ in detail.  This serves to illustrate features that
we  take over to $\mathcal{N}=8$, thus obtaining the linearized
$\mathcal{N}=8$ super-NMG theory. Finally, we  construct the
off-shell $\mathcal{N}=7$ LCS invariant and hence the
maximally-supersymmetry parity-violating super-GMG and super-TMG
theories, again at the linearized level. We conclude with a
discussion of open problems.


\section{$\mathcal{N}$-extended Weyl and Poincar\'{e} supermultiplets}

To construct the $\mathcal{N}$-extended supersymmetric  LCS term,
one needs only the fields of an $\mathcal{N}$-extended  ``Weyl
supermultiplet''. At the  linearized level, this multiplet contains
the metric perturbation $h_{\mu\nu}$ and $\mathcal{N}$ Majorana
anti-commuting vector-spinors $\psi_\mu^i$ ($i=1,
\dots,\mathcal{N}$) subject to the linearized transformations
\be\label{lineargauge1} \delta h_{\mu\nu} = \partial_{(\mu} v_{\nu)}
+ \eta_{\mu\nu} \omega\, , \qquad \delta \psi_\mu^i = \partial_\mu
\eta_Q + \gamma_\mu\epsilon_S\, , \ee for Minkowski 3-vector $v$,
scalar $\omega$ and the anticommuting Majorana spinor parameters of
$Q$- and $S$-supersymmetry.  At the linearized level we must
distinguish between the $Q$-supersymmetry gauge invariances and the
rigid supersymmetry that relates the various fields of the
multiplet; it is only after the inclusion of interactions that these
combine to become $\mathcal{N}$ local supersymmetries.  The
linearized Weyl multiplet also includes
$\mathcal{N}(\mathcal{N}-1)/2$  abelian gauge fields $V_\mu^{ij}$ in
the adjoint  irrep of the $Spin({\cal N})$ R-symmetry group, and
subject to the gauge transformation \be \label{lineargauge2} \delta
V_\mu^{ij} = \partial_\mu \Lambda^{ij}\, . \ee For $\mathcal{N}\ge3$
there are additional fields in the Weyl multiplet, and for
$\mathcal{N}\ge6$ these include additional gauge fields, although
$\mathcal{N}=8$ is exceptional in this respect, as we shall see.

To construct the $\mathcal{N}$-extended supersymmetric EH term, or
$\mathcal{N}$-extended ``Einstein supergravity'', we need more than
just the Weyl supermultiplet. We need a Poincar\'e supermultiplet in
which the fields are subject only to the standard supergravity gauge
transformations and not the additional ones of conformal
supergravity. We therefore need to introduce additional degrees of
freedom. One way to do this is as follows
\cite{Siegel:1977hn,Kaku:1978ea}. We  start from a standard flat
space action invariant under rigid superconformal transformations,
for a supermultiplet (or supermultiplets) containing physical scalar
and spinor fields (and possibly vector or antisymmetric tensor gauge
fields);  if  the field supermultiplets used to construct this
action are off-shell supersymmetric then there will generically be
auxiliary fields too but their inclusion is optional.   This action
is then  coupled to an off-shell $\mathcal{N}$-extended Weyl
supermultiplet (e.g. by the Noether procedure) to give an action
that is invariant under local superconformal transformations.  Then
one fixes the `unwanted' superconformal symmetries by imposing
conditions on the physical scalar and spinor fields of the
multiplet(s).  To do so, one needs sufficient multiplets so as to
have a sufficient number of physical scalar and spinor fields, and
with luck there will be none left over. In this case, the conformal
coupling to the scalar fields will produce the Einstein-Hilbert term
on gauge fixing and the rest of the action will be just what is
required  for the $\mathcal{N}$-extended supersymmetrization of this
term.  One says that the original `scalar' supermultiplets are
`compensating multiplets'. The auxiliary fields of these
compensating multiplets become auxiliary fields of the  final
$\mathcal{N}$-extended Einstein supergravity action.

If we wish to construct generic higher-derivative actions in the way
just described then  the auxiliary fields of the compensating
multiplets must be included because  the starting flat space action
for the conformal compensator multiplets must be a higher-derivative
one in which the `auxiliary'  fields also propagate. The LCS term is
an exception to this rule because it is superconformally invariant.
In contrast, the 4th order curvature-squared term of NMG is not
conformally invariant, so one really needs a full superconformal
tensor calculus to construct $\mathcal{N}$-extended NMG.  Some
aspects of this calculus have been worked out for $\mathcal{N}=1,2$
\cite{Uematsu:1984zy} but, even so, special `tricks'  were needed
for the construction of the $\mathcal{N}=1$ supersymmetric super-NMG
model \cite{Andringa:2009yc,Bergshoeff:2010mf} and only linearized
results have been found for $\mathcal{N}=2$ \cite{Andringa:2009yc}.

However, although we need an $\mathcal{N}$-extended superconformal
tensor calculus to construct the full $\mathcal{N}$-extended NMG
action, we do not need it to construct the quadratic approximation
to this action. This is because the NMG 4th order invariant has the
property (shared with an infinite series of yet higher-order
invariants) that its quadratic approximation is invariant under the
linearized super-Weyl gauge invariances, so compensating multiplets
are not needed. Moreover, this quadratic approximation to the 4th
order NMG invariant may be added to the quadratic approximation to
the $\mathcal{N}$-extended Einstein supergravity to give the
quadratic approximation to $\mathcal{N}$-extended NMG.  The upshot
is that to construct the $\mathcal{N}$-extension of linearized NMG,
and more  generally the linearized GMG, we do not need a full
superconformal tensor calculus. In particular, we need conformal
compensator supermultiplets only for  the construction of the
$\mathcal{N}$-extended Einstein supergravity, and as we need that
only to quadratic level, it is sufficient to consider the cubic
interaction of the compensator supermultiplets to the Weyl multiplet
and a few other quartic interactions that contribute to the
quadratic action after fixing the superconformal gauge.

In what follows we will consider sequentially the cases from
$\mathcal{N}=2,\dots,8$, presenting some details of the Weyl
multiplet and determining the type and number of compensating
multiplets that are needed. For the convenience of the reader we
have  summarized some details about this in Table 2. The
$\mathcal{N}=2$ case has already been dealt with in some detail
\cite{Howe:1995zm,Andringa:2009yc} but it will serve to illustrate
the issues involved. \vskip .1truecm

\begin{table}[h]
\caption{{\footnotesize{Some properties of the ${\cal N}$--extended
Weyl multiplets. The fourth column indicates the number of basic
compensating supermultiplets needed to obtain (Einstein, TMG or GMG)
supergravity. }}} \vspace{0.1cm} \centering
\begin{tabular}{cccccccccc}
\hline\hline\\[-.2truecm]
${\cal N}$&\# off-shell d.o.f.&R-symmetry&\# multiplets\\[.2truecm]
\hline\\[-.2truecm]
2&4+4&SO(2)&1\\[.2truecm]
3&8+8&SO(3)&1\\[.2truecm]
4&16+16&$Spin(4)\cong SU(2)\times SU(2)$&2\\[.2truecm]
5&32+32&$Spin(5)\cong Sp_2$&2\\[.2truecm]
6&64+64&$Spin(6)\cong SU(4)$&4\\[.2truecm]
7&128+128&$Spin(7)$&8\\[.2truecm]
8&128+128&SO(8)&8\\[.2truecm]
\hline
\end{tabular}
\end{table}

\vskip .2truecm


\subsection{$\mathcal{N}=2$}

The off-shell linearized Weyl multiplet has the field content \be
(h_{\mu\nu}; \psi_\mu^i; V_\mu) \, , \qquad i=1,2. \ee We use
semicolons to separate fields of different mass dimensions, which
increase by steps of 1/2.  The vector field is the $SO(2)$ gauge
field. Taking all gauge invariances into account, we are left with
$4+4$ remaining off-shell field components.

To construct a Poincar\'e supermultiplet suitable for the
construction of massive supergravities, we need to add additional
degrees of freedom that will allow us to fix  the scale and
$S$-supersymmetry transformations, at least, but we may also fix the
R-symmetry gauge invariance. For example, consider the
$\mathcal{N}=2$ scalar multiplet with $4+4$ off-shell field content
$(\varphi, \varsigma; \lambda^i ; S,P)$. After coupling to the Weyl
multiplet, we may fix the scale and $SO(2)$ gauge invariances by
setting $\varphi=0$ and $\varsigma=0$, while the $S$-supersymmetry
gauge invariances may be fixed by setting $\lambda^i=0$. This leaves
the auxiliary fields $S$ and $P$, which survive as auxiliary fields
of the  off-shell Poincar\'e supermultiplet, which has the $8+8$
off-shell field content $(h_{\mu\nu}; \psi_\mu^i; V_\mu, S,P)$,  now
subject only to diffeomorphisms and local  $Q$-supersymmetry gauge
transformations.  This multiplet was called  the $(1,1)$ Poincar\'e
supermultiplet in  \cite{Howe:1995zm} in order to distinguish it
from the $(2,0)$ Poincar\'e supermultiplet found by taking the
compensator fields to belong to the $4+4$ off-shell vector
multiplet, which has the field content $(\varphi,  A_\mu; \lambda^i
; D)$. In this case, we use $\varphi$ and $\lambda^i$ to compensate
for the scale and $S$-supersymmetry transformations, as before, but
we leave uncompensated the local $SO(2)$ gauge invariance. The
linearized Poincar\'e supermultiplet now has the $8+8$ off-shell
field content $( h_{\mu\nu}; A_\mu; \psi_\mu^i; V_\mu; D)$. In the
construction of the supersymmetric Einstein-Hilbert term the $(2,0)$
Poincar\'e supermultiplet leads to a $VdA$ Chern-Simons type term
\cite{Howe:1995zm}, and a $D^2$ term for the auxiliary field $D$.

Variant choices of conformal compensator multiplets therefore lead
to variant versions of the supersymmetric EH action; in  this case
the $(1,1)$ or  $(2,0)$ versions.   Whatever the choice, one can add
to this action any off-shell supersymmetric $\mathcal{N}=2$
superconformal action, such as LCS. Also, at the linearized level,
we can add the quadratic approximation to  the  $\mathcal{N}=2$
supersymmetric 4th order invariant of NMG, thereby constructing the
$\mathcal{N}=2$  extension of  linearized GMG. This was done for the
$(1,1)$ case in \cite{Andringa:2009yc}.  The generic GMG model
propagates one multiplet of helicities  $(2,3/2,1)$ with mass $m_+$
and another multiplet of helicities  $(-1,-3/2,-2)$ with mass $m_-$.
The same computations may be done for the $(2,0)$ case, but with
more difficulty because the zero Weyl weight of the vector field $A$
means that the starting flat space action cannot be quadratic.
Nevertheless, it can be done and the final spectrum is the same but
the spin 1 modes are propagated differently. In the $(1,1)$ case the
action for the spin 1 modes is  the Proca  action  for $V$ with a CS
term\footnote{But here we use a different notation: what was called
$A$   in \cite{Andringa:2009yc} is $V$ here.}. In the $(2,0)$ case
one finds that the action for the spin 1 modes has the Lagrangian
\be L_1 = \frac{1}{2}\tilde F^2 +  A^\mu \tilde G_\mu +
\frac{1}{2\mu} V^\mu \tilde G_\mu  +  \frac{1}{2m^2} \tilde G^2 \ee
where \be \tilde F^\mu = \varepsilon^{\mu\nu\rho} \partial_\nu
A_\rho\, , \qquad \tilde G^\mu = \varepsilon^{\mu\nu\rho}
\partial_\nu V_\rho\, . \ee One may verify that this propagates two
spin 1 modes of helicities $\pm1$ and masses $m_\pm$ given by \be
m^2 = m_+ m_- \, \qquad \mu = m^2/(m_--m_+)\, , \ee exactly as for
the $(1,1)$ case.


\subsection{$\mathcal{N}=3$}

The off-shell $\mathcal{N}=3$ linearized Weyl multiplet has $8+8$
components and the field content \be (h_{\mu\nu}; \psi^i_\mu;
V_\mu^i; \chi)\, ,  \qquad i=1,2,3. \ee The minimal $\mathcal{N}=3$
scalar multiplet has the physical (i.e. non-auxiliary) field content
$(\varphi, \varphi^i; \lambda, \lambda^i)$; i.e.~both scalars and
spinors are in the $1 \oplus 3$ representation of the
$\mathrm{SO}(3)$ $R$-symmetry group. There is an off-shell version
of this multiplet  with a finite number of auxiliary fields (a 3D
version of the relaxed hypermultiplet \cite{Howe:1982tm}) but it is
not needed for present purposes.  The $1 \oplus 3$ scalars are
precisely those needed to compensate for the one scale and three
$\mathrm{SO}(3)$ gauge invariances. The fermion triplet $\lambda^i$
compensates for the three $S$-supersymmetries. This leaves the
singlet spinor $\lambda$. In the context of the $\mathcal{N}=3$
supersymmetrization of the Einstein-Hilbert action, the pair of
spinors $(\chi,\lambda)$  is auxiliary. As the (non-gauge) vectors
$V_\mu^{ij}$ are also auxiliary in this context, only $h_{\mu\nu}$
and $\psi_\mu^i$ remain as the 'physical' fields,
 which actually do not propagate modes.

In the context of $\mathcal{N}=3$ super-TMG there is a $\bar{\chi}
\chi$ term associated with the Lorentz-Chern-Simons term and this
means that $\chi$ is no longer a Lagrange multiplier for the
constraint $\lambda=0$. The combined effect of the `$\chi \lambda$'
terms is to propagate one spin-$1/2$ mode. Thus, the physical field
content in the context of TMG is $(h_{\mu\nu};  \psi^i_\mu , \lambda
;  V_\mu^i; \chi)$ subject only to linearized diffeomorphism and
local $Q$-supersymmetry transformations.  Note that the effect of
the compensating multiplet is no longer  merely to reduce the gauge
invariances of the Weyl multiplet fields, as is the case for
$\mathcal{N}=2$, but also to provide an additional spinor field. The
physical field content for NMG is the same as for TMG but we expect
the higher-order interactions to lead to a second massive particle
supermultiplet of opposite helicities in exactly the way spelled out
for $\mathcal{N}=2$ in \cite{Andringa:2009yc}.

As for $\mathcal{N}=2$, we could consider how things change if we
use vector multiplets in place of scalar multiplets, but this is
already much more complicated for $\mathcal{N}=3$ so we shall
henceforth restrict the discussion to scalar supermultiplets.


\subsection{$\mathcal{N}=4$}

The  off-shell  $\mathcal{N}=4$ Weyl multiplet, with 16+16
components,  has the field content \be\label{n=4weyl} (h_{\mu\nu};
\psi^i_\mu; V_\mu^{ij}, E; \chi^i ; D)\, , \qquad i=1,2,3,4. \ee The
R-symmetry group is now reducible:  $Spin(4)\cong SU(2)\times
SU(2)$. As a consequence, the vector fields may be written as \be
V^{ij}= V_+^{ij} + V_-^{ij}\, , \qquad \frac{1}{2}\varepsilon^{ijkl}
V_\pm ^{kl} = \pm V_\pm^{kl}\, . \ee This sum of self-dual and
anti-self-dual terms corresponds to the direct sum representation
$(3,1)\oplus (1,3)$. This sum is of course unchanged under a switch
in the duality assignments of $V_\pm$, which amounts to an exchange
of the two $SU(2)$ factors of $Spin(4)$ by an outer $\mathbb{Z}_2$
automorphism.  However, the scalars $E,D$ and the spinors $\chi^i$
are {\it odd} under this $\mathbb{Z}_2$ exchange\footnote{It was
suggested in \cite{Howe:1995zm} that there should exist an 8+8 Weyl
multiplet that contains only $V_+$ or $V_-$, and without  the higher
dimension fields $(E;\chi^i;D)$, but it does not appear to be
possible to close the supersymmetry algebra on this smaller set.}.

The minimal scalar ${\cal N}=4$ supermultiplet is the 3D analog of
the 4D, $\mathcal{N}=2$ hypermultiplet with $4+4$ physical degrees
of freedom. However, in 3D there are two distinct versions of the
hypermultiplet. The hypermultiplet scalars transform as a complex
doublet of one or the other of the $SU(2)$ factors of the R-symmetry
group. Whichever $SU(2)$ factor we choose,  the spinor fields  will
transform as a complex doublet of the other $SU(2)$ factor.  In
other words, we have a hypermultiplet with physical fields
$(\varphi^\alpha,\lambda^{\dot\alpha})$ or one with physical fields
$(\varphi^{\dot\alpha},\lambda^{\alpha})$, where $\alpha=1,2$ and
$\dot\alpha=1,2$ are spinor indices for the two $SU(2)$ factors of
$Spin(4)$. The $SU(2)$ currents that may be constructed from these
scalars will be either self-dual or anti-self-dual, according to
which of the two versions of the hypermultiplet we choose.  Without
loss of generality, we may assume that  the scalar fields
$\varphi^\alpha$ yield a self-dual current $J_+^{ij}$ and that the
scalar fields  $\varphi^{\dot \alpha}$ yield an anti-self-dual
current  current $J_-^{ij}$.

Both types of hypermultiplet are needed for conformal compensation,
because we have to compensate for both $\mathrm{SU}(2)$ gauge
invariances. Another difference from the $\mathcal{N}=3$ case is
that any $R$-symmetry singlet constructed from the scalar fields of
the compensating multiplets must be at least quadratic in the scalar
fields. The scalar bilinears \be\label{defs} \varphi^2_+ \equiv
\varphi^\alpha \varphi_\alpha + \varphi^{\dot{\alpha}}
\varphi_{\dot{\alpha}}\, , \qquad \varphi^2_- \equiv  \varphi^\alpha
\varphi_\alpha - \varphi^{\dot{\alpha}} \varphi_{\dot{\alpha}} \ee
are respectively even and odd under the $\mathbb{Z}_2$ exchange. The
former acts as a compensator for the dilatations. After fixing the
dilatation gauge by imposing $\varphi^2_+=1$, one finds that the
$Spin(4) $ currents take the form \be\label{goldstone}
J_\mu^{ij}{}_\pm = \partial_\mu \varphi_\pm ^{ij} + \dots\;, \ee
where $\varphi_\pm ^{ij}$ are, collectively,  $Spin(4)$ Stueckelberg
scalars  that may be used to compensate for the local $Spin(4)$
gauge invariance.

In contrast to $\mathcal{N}=3$, the conformal compensation mechanism
is closer to the non-linear Higgs mechanism than it is to the linear
Stueckelberg mechanism, because $\varphi^2_-$ can be identified as
the field of a residual Higgs boson that survives the `spontaneous'
breaking of symmetries implied by the conformal gauge condition
$\varphi_+^2=1$.  However, this Higgs boson field is set to zero by
the $D$ field equation of the Weyl multiplet.  The details of the
construction will be discussed in the next section.


\subsection{$\mathcal{N}=5$}

The off-shell linearized Weyl multiplet  for $\mathcal{N}=5$ has
$32+32$ components and the field content \be (h_{\mu\nu};
\psi^i_\mu; V_\mu^{ij}, E^i; \chi^{ij}, \chi; D^i)\, ,  \qquad
i=1,2,3,4,5. \ee Fields with multiple indices are antisymmetric in
these indices. Thus, the vectors are in the adjoint ${\bf 10}$ irrep
of $Spin(5)\cong Sp_2$, as are the spinors $\chi^{ij}$.

The minimal scalar supermultiplet now has $8 + 8$ physical fields,
in the complex $4$ representation of $\mathrm{Sp}_2$.  Let
$(\varphi^\alpha, \lambda^\alpha)$ $(\alpha = 1, 2, 3,4)$ be these
physical fields.  As for $\mathcal{N}=4$, we  need {\it two} scalar
supermultiplets, $\varphi_1$ and $\varphi_2$.  The 16  scalar fields
in these two multiplets may be traded for  scalar bilinears in the
${\bf 1}\oplus {\bf 5}$ representations of  $Sp_2$, and currents in
the adjoint ${\bf 10}$ of $Sp_2$.  The singlet bilinear is
$\Omega_{\alpha\beta} \varphi^\alpha_1 \varphi^\beta_2$, which we
set to unity to fix the local scale invariance.  The currents then
become derivatives of a ${\bf 10}$ of Stueckelberg scalars, to
linear order, which may be set to zero to compensate for the local
$Sp_2$ invariance. The  5-plet of scalar bilinears is set to zero by
the $D^i$ field equation.


\subsection{$\mathcal{N}=6$}

The off-shell linearized Weyl multiplet now has $64+64$ components,
and the field content \be (h_{\mu\nu}; \psi^i_\mu; V_\mu^{ij},
V_\mu, E^{ij}; \chi^{ijk}, \chi^i; D^{ij}) \, ,  \qquad
i=1,2,3,4,5,6. \ee Fields with multiple indices are again
antisymmetric in these indices. The $R$-symmetry group is $Spin(6)
\cong \mathrm{SU}(4)$, but the gauge symmetry is enhanced to
$\mathrm{U}(4)$, because of the presence of the additional gauge
field $V_\mu$.

The minimal scalar multiplet again has $8 + 8$ physical fields
$(\varphi^\alpha, \chi^\alpha)$ in the (complex) ${\bf 4}$  of
$\mathrm{SU}(4)$. We can choose a gauge given by the sum of squares
of all the scalars equal to one to fix the dilatations. There will
be 15 additional constraints coming from the $D$-field equation. The
$\mathrm{U}(4)$ currents then become to linearized order the
derivatives of 16 Stueckelberg scalars, which we set to zero to fix
the $\mathrm{U}(4)$ invariance. We therefore need a total of 32
scalar fields and hence {\it four} scalar multiplets.


\subsection{$\mathcal{N}=7$}

The off-shell linearized Weyl multiplet has $128+128$ components,
and the field content \be (h_{\mu\nu}; \psi^i_\mu, \psi_\mu;
V_\mu^{ij}, V_\mu^i, E^{ijk}; \chi^{ijk}, \chi^{ij}; D^{ijk}) \, ,
\qquad i=1,2,3,4,5,6,7. \ee Fields with multiple indices are again
antisymmetric in these indices. The $R$-symmetry group is $Spin(7)$
but the presence of 7 additional vector fields implies an
enhancement of the gauge symmetry. We also have an additional
Rarita-Schwinger field implying an enhanced local supersymmetry at
the non-linear level, but we avoid the associated difficulties here
by restricting to the linear theory.

The minimal scalar multiplet has 8 + 8 components in the spinor
representation of $\mathrm{Spin}(7)$. The $D$-fields now impose 35
constraints on scalar bilinears. In addition, we need $1 + 21 + 7 =
29$ scalars to compensate for the scale and gauge invariances. In
total we need 64 scalars and hence {\it eight} compensating scalar
multiplets.


\subsection{$\mathcal{N}=8$}

The off-shell Weyl multiplet, again with 128+128 components, has the
field content \be \label{weylmultn8} (h_{\mu\nu}\ ; \ \psi^i_\mu \ ;
\ V_\mu^{ij}, E^{ijkl} \ ; \ \chi^{ijk} \ ; \ D^{ijkl}) \, ,  \qquad
i=1,2,3,4,5,6,7,8. \ee Fields with multiple indices are again
antisymmetric in these indices, and  $E^{ijkl}$ and $D^{ijkl}$ have
'opposite $SO(8)$ dualities': they are, respectively, self-dual and
anti-selfdual,  or vice-versa. As there are two choices of duality
assignments, there are two (equivalent) versions of the
$\mathcal{N}=8$ Weyl multiplet.

The minimal scalar multiplet has 8 + 8 physical components and also comes in two versions. One version has the scalars in the spinor representation and the fermions in the conjugate spinor representation of $\mathrm{Spin}(8)$ and vice versa for the other version. These are the only possibilities consistent with the supersymmetry parameter being a vector of $\mathrm{SO}(8)$. However, only one of the two versions of the scalar multiplet can be consistently coupled to a given version of the Weyl multiplet, and hence all compensating supermultiplets must be of the same type. The $D$-field will impose 35 constraints on scalar bilinears and we need $1 + 28$ scalars to compensate for the scale and $\mathrm{SO}(8)$ gauge invariances. We therefore need a total of 64 scalars and hence {\it eight}  compensating scalar multiplets.


\section{$\mathcal{N}=4$ Massive Supergravities}
\label{sec:N=4}

In this section we present details of the construction of linearized
3D $\mathcal{N}=4$ supergravities with various higher-derivative
interactions using the superconformal tensor calculus. In
particular, we obtain the $\mathcal{N}=4$ extension of linearized
GMG and its limits. Essential to all these constructions is the
$16+16$ component linearized Weyl multiplet of (\ref{n=4weyl}). The
linearized supersymmetry transformation rules (in the same
conventions as in \cite{Andringa:2009yc}) are
\bea \label{linsusyn4} \delta h_{\mu\nu} &=& {\bar\epsilon}^i\gamma_{(\mu}\psi_{\nu)}^i\,,\nonumber\\[.2truecm]
\delta\psi_\mu^i &=& -\tfrac{1}{4}\gamma^{\rho\sigma}\partial_\rho
h_{\mu\sigma}\epsilon^i
- V_\mu^{ij}\epsilon^j \,,\nonumber  \\[.2truecm]
\delta V_\mu^{ij} &=& \tfrac{1}{2} {\bar\epsilon}^{[i}\phi_\mu^{j]}   +  \varepsilon^{ijkl} \bar{\epsilon}^k \gamma_\mu \chi^l \,, \nonumber \\[.2truecm]
\delta E &=&  \tfrac{1}{4} \bar{\epsilon}^i \chi^i \,,   \\[.2truecm]
\delta\chi^{i} &=& \tfrac{1}{8} \varepsilon^{ijkl} \gamma^\mu \epsilon^l F_{\mu\,\text{(lin)}}^{jk} +  \gamma^\mu \partial_\mu E\epsilon^i  + D \epsilon^i\,, \nonumber \\[.2truecm]
\delta D &=& \tfrac{1}{4} \bar{\epsilon}^i \gamma^\mu
\partial_\mu \chi^i \,. \nonumber \eea
Here $F^{\mu\,ij}_{\rm (lin)}$ and the (dependent) $S$-supersymmetry
gauge field $\phi_\mu^i$ are given by
\begin{equation} \label{defFphi}
F^{\mu\,ij}_{\rm (lin)} = \epsilon^{\mu\nu\rho}\partial_\nu
V^{ij}_\rho\,, \quad \phi_\mu^i = \gamma_\nu\gamma_\mu {\cal R}_{\rm
(lin)}^{\nu\,i}\,, \quad {\cal R}_{\rm (lin)}^{\mu\,i} =
\epsilon^{\mu\nu\rho}\partial_\nu\psi_\rho^i\,.
\end{equation}
We furthermore insist on gauge invariance with respect to the
linearized gauge transformations of (\ref{lineargauge1}) and
(\ref{lineargauge2}). The supersymmetry algebra now closes in the
sense that the commutator of two supersymmetries on any field, with
parameters $\epsilon_1$ and $\epsilon_2$, gives a translation
 plus (field-dependent) gauge transformations (represented by the dots):
\be
[\delta(\epsilon_1),\delta(\epsilon_2)] = \frac{1}{2}\left({\bar\epsilon}^i_2\gamma^\mu\epsilon^i_1\right)\partial_\mu
+ \cdots\,  \, .
\ee

\subsection{Further properties of the Weyl multiplet fields}

The transformation laws given above for the independent fields of
the Weyl multiplet imply the  following supersymmetry transformation
of the dependent $S$-supersymmetry gauge field: \be \delta
\phi_\mu^i = \gamma^\nu \epsilon^i S^{(lin)}_{\mu\nu} -
\gamma_\nu\gamma_\mu \epsilon^j F_{(lin)}^{\nu\,  ij}\, , \ee where
$S_{\mu\nu}$ is the linearized 3D Schouten tensor \be
S^{(lin)}_{\mu\nu} = R_{\mu\nu}^{(lin)} - \frac{1}{4} \eta_{\mu\nu}
R^{(lin)}\, , \qquad
 R^{\rm (lin)} \equiv  \eta^{\mu\nu}R_{\mu\nu}^{\rm (lin)}\, .
\ee
Using the expression
\begin{equation}
R_{\mu\nu}^{\rm (lin)} = -\tfrac{1}{2}\big[\Box h_{\mu\nu} -
2\partial_{(\mu}h_{\nu)}+\partial_\mu\partial_\nu h\big]\,, \qquad h_\mu \equiv \eta^{\nu\rho}\partial_\rho h_{\mu\nu}\, , \quad
h = \eta^{\mu\nu}h_{\mu\nu}\,,
\end{equation}
one can verify that the linearized Weyl transformation of the
linearized Schouten tensor takes the following simple form \be
\delta_\omega S^{(lin)}_{\mu\nu} = -\frac{1}{2}
\partial_\mu\partial_\nu \omega\, . \ee It follows that the
linearized Cotton tensor, defined as \be C_{\mu\nu}^{(lin)} =
\varepsilon_\mu{}^{\tau\rho} \partial_\tau S^{(lin)}_{\rho\nu}\, ,
\ee is linearized Weyl invariant; this is a consequence  of the
Weyl invariance of the non-linear Cotton tensor, which is the 3D
analog of the 4D Weyl tensor. The linearized Cotton tensor, which is
parity odd,  satisfies the identities \be
\partial^\mu C_{\mu\nu}^{(lin)} \equiv 0\, , \qquad C^{(lin)}_{\mu\nu} \equiv C^{(lin)}_{\nu\mu} \, , \qquad
\eta^{\mu\nu}C_{\mu\nu}^{(lin)} \equiv 0\, ,
\ee
which are also consequences of similar identites satisfied by the full Cotton tensor.

The superpartner of the Cotton tensor is the Cottino tensor.   The linearized $\mathcal{N}=4$ Cottino tensor is
\be
{\cal C}_{\rm (lin)}^{\mu\, i} =  \gamma^\nu\partial_\nu {\cal
R}^{\mu\,i}_{\rm (lin)} + \epsilon^{\mu\nu\rho}\partial_\nu{\cal
R}^i_{\rho\,{\rm (lin)}}\,.
\ee
It satisfies the identities
\begin{equation} \label{cottinoid}
\partial_\mu{\cal C}^{\mu\,i}_{\rm (lin)}=0\,,\qquad \gamma_\mu{\cal C}^{\mu\,i}_{\rm (lin)}=0\,.
\end{equation}
The linearized Cotton, Cottino and $\text{SO(4)}$ curvature tensors have the following supersymmetry transformations:
\begin{eqnarray}\label{cotttrans}
\delta C_{\mu\nu}^{\rm (lin)} &=&  -\tfrac{1}{4}{\bar\epsilon}^i\gamma_{(\mu}{}^\rho\partial_\rho{\cal C}_{\nu)}^{i\,{\rm lin}}\,,   \nonumber\\[.2truecm]
\delta{\cal C}^{\mu\, i}_{\rm (lin)} &=&  \gamma_\nu\epsilon^i
C^{\mu\nu}_{\rm (lin)}
+\epsilon^{\mu\nu\rho}\gamma_\sigma\gamma_\nu\epsilon^j\partial_\rho F^{\sigma\,ij}_{\rm (lin)}\,,  \\[.2truecm]
\delta F^{\mu\,ij}_{\rm (lin)} &=&
\tfrac{1}{2}{\bar\epsilon}^{[i}{\cal C}^{\mu\, j]}_{\rm (lin)} +
{\bar\epsilon}^k\epsilon^{ijkl}\gamma^{\mu\rho}\partial_\rho\chi^{l}\;.
\nonumber
\end{eqnarray}
These transformation rules define an ${\cal N}=4$ conformal  field-strength  multiplet with components
\begin{equation}\label{Weylc4}
\big \{C_{\mu\nu}^{\rm (lin)}\,,{\cal C}^{\mu\, i}_{\rm
(lin)}\,,F^{\mu\,ij}_{\rm (lin)}\,,\chi^{i}\,,D\,, E
\big\}\,.
\end{equation}

With the above definitions in hand we may construct various
conformal higher-derivative actions for the conformal multiplet.
Below we give a few examples of such actions where the leading term,
bilinear in the graviton field, has 3,4,5 and 6 derivatives.

\subsection{${\cal N}=4$ linearized Weyl multiplet actions}

Let us now consider invariants that may be constructed from the Weyl multiplet fields alone, at least in the quadratic approximation.

\bigskip

\noindent (1)\ $\mathcal{N}=4$ {\sl Supersymmetric LCS}\hfill\break
\vskip -.4truecm
One may verify that the following action containing a linearized Lorentz Chern-Simons term is
supersymmetric,
\begin{equation}\label{linactionN=4}
S_3^{{\cal N}=4} = \int d^3x\,\bigg\{h^{\mu\nu}C_{\mu\nu}^{\rm
(lin)} + {\bar\psi}^i_\mu{\cal C}_{\rm (lin)}^{\mu\,i} -2
V_\mu^{ij}F_{\rm (lin)}^{\mu\,ij} +16{\bar\chi}^{i}\chi^{i}-128
ED\bigg\}\,.
\end{equation}
As this action is the quadratic approximation to an $\mathcal{N}=4$
superconformal extension of the LCS term, no compensating fields are
needed to construct the non-linear invariant.

\bigskip

\noindent (2)\ $\mathcal{N}=4$ {\sl Supersymmetric} NMG invariant \hfill\break
\vskip -.2truecm

Using (\ref{cotttrans}), one may verify  invariance of the following action:
\begin{eqnarray}\label{KactionN=4}
S_4^{{\cal N}=4} &=& \int
d^3x\,\bigg\{-\tfrac{1}{2}\epsilon^{\mu\tau\rho}h_\mu{}^\nu\partial_\tau
C_{\rho\nu}^{\rm (lin)} -\frac{1}{2} {\bar\psi}^i_\mu \partial
\hskip -.24truecm  / {\cal C}^{{i\, \mu\, \rm (lin)}}
+  F_{\mu\,{\rm (lin)}}^{ij} F_{\rm (lin)}^{\mu\,ij} +\nonumber\\[.2truecm]
&&\hskip 4truecm +32 E\Box E -8 {\bar\chi}^{i}\partial \hskip
-.24truecm  /\,\chi^{i}+32 D^2\bigg\}\,.
\end{eqnarray}
The leading term is just the linearization of the fourth-order $K$ invariant of NMG \cite{Bergshoeff:2009hq}. To see this
we note first that  $K\equiv R^{\mu\nu}R_{\mu\nu} - \frac{3}{8} R^2 = G^{\mu\nu}S_{\mu\nu}$.
A convenient form for the linearized 3D Einstein tensor is
\be
G^{\mu\nu}_{(lin)} = -\frac{1}{2} \varepsilon^{\mu\tau\rho} \varepsilon^{\nu\eta\sigma}\partial_\tau\partial_\eta h_{\rho\sigma}\;.
\ee
Using this, one may show that
\be\label{altNMG}
G_{(lin)}^{\mu\nu}S_{\mu\nu}^{(lin)} = -\frac{1}{2} \varepsilon^{\mu\tau\rho} h_\mu{}^\nu \partial_\tau C_{\rho\nu}^{(lin)} \
+ \  {\rm total\ derivative}\;.
\ee
The expression on the right hand side makes manifest the linearized Weyl invariance.

\bigskip

\noindent (3)\ ${\cal N}=4$ {\sl Supersymmetric Ricci times Cotton}\hfill\break
\vskip -.2truecm

One can also construct a fifth-order parity-odd  action that starts
with the product of a Ricci tensor with a Cotton tensor:
\begin{eqnarray}\label{5thorder}
S_5^{{\cal N}=4} &=& \int d^3x\,\bigg\{R^{\mu\nu}_{\text{(lin)}} C_{\mu\nu}^{\rm (lin)} + {\bar {\cal C}}_{\mu\, i}^{\text{(lin)}} {\cal C}_{\rm (lin)}^{\mu\,i}+\epsilon^{\mu\nu\rho}F_\mu^{\text{(lin)}\, ij}\partial_\nu F^{\text{(lin)}\, ij}_{\rho} \nonumber\\[.2truecm]
&&\hskip 2.5truecm +16{\bar\chi}^{i}\Box \chi^{i}-128 E\Box
D\bigg\}\,.
\end{eqnarray}

\bigskip

\noindent (4)\ ${\cal N}=4$ {\sl Supersymmetric Cotton Tensor Squared}\hfill\break
\vskip -.2truecm

It is not difficult to construct linearized higher-derivative supersymmetric actions  by using the following
observation. Schematically the supersymmetric action  (\ref{linactionN=4}) consists
of terms that are all of the form
\begin{equation}\label{firstaction}
\text{Weyl} \times \text{Weyl}^c\,,
\end{equation}
where Weyl indicates a field from the Weyl multiplet
and $\text{Weyl}^c$ denotes a field from the
conformal field-strength multiplet \eqref{Weylc4}.
The fact that the supersymmetry transformations are linearized and thus global
has the following immediate consequence:
Given a set of fields $\{\text{Weyl}\}$ that transforms as
\eqref{linsusyn4}, the set of fields
\begin{equation}
\{\text{Weyl}^{(n)}\} = \{\Box^n\text{Weyl}\}
\end{equation}
transforms in the same way. This means that the  action that is
obtained from \eqref{KactionN=4} by replacing in each term the
first factor Weyl by $\text{Weyl}^{(n)}$, such that we obtain
\begin{equation}\label{secondtaction}
\text{Weyl}^{(n)} \times \text{Weyl}^c\,,
\end{equation}
also defines a linearized supersymmetric invariant. The case $n=1$
is particularly interesting since one can rewrite, using conformal
gauge transformations, the action such that each term is bilinear in
the fields of the conformal field-strength Weyl multiplet. This leads to the
supersymmetric Cotton tensor squared action:
\begin{eqnarray}\label{CactionN=4}
S_6^{{\cal N}=4} &=& \int d^3x\,\bigg\{C_{\rm
(lin)}^{\mu\nu}C_{\mu\nu}^{\rm (lin)} -\frac{1}{4} {\bar {\cal
C}}^{i\,{\rm (lin)}}_\mu  \partial \hskip -.24truecm  / {\cal
C}_{\rm (lin)}^{\mu\,i}
+  F_{\mu\,{\rm (lin)}}^{ij}\Box F_{\rm (lin)}^{\mu\,ij} +\nonumber\\[.2truecm]
&&\hskip 2truecm +32 E\,\Box^2 E -8 {\bar\chi}^{i}\Box\,\partial
\hskip -.24truecm  /\,\chi^{i}+32 D\,\Box D\bigg\}\,.
\end{eqnarray}
%

This concludes our discussion of linearized  higher-derivative
actions that can be constructed from the fields of the Weyl
multiplet alone.


\subsection{${\cal N}=4$  Einstein supergravity}

To obtain the ${\cal N}=4$ supersymmetric Einstein action we need two  compensating hypermultiplets, one of each type, for reasons explained in the previous section.  Following the superconformal approach we couple these compensating hypermultiplets to the Weyl multiplet and  take suitable gauge
choices for the superfluous (super-) conformal symmetries. This general procedure simplifies due to the fact that we only consider the linearized version
of ${\cal N}=4$ Einstein supergravity.

The hypermultiplet  with physical  fields $(\varphi^{\dot{\alpha}}, \lambda_{\alpha})$ has the  supersymmetry transformations
\be \label{susyhyper} \delta \varphi^{\dot{\alpha}} =
\bar{\epsilon}^i
\lambda_\beta \bar{\sigma}^{i\, \dot{\alpha}\beta} \,, \qquad
\delta \lambda_\alpha =  \tfrac{1}{4} \gamma^\mu \partial_\mu
\varphi^{\dot{\beta}} \epsilon^i \sigma^i_{\alpha \dot{\beta}} \,,
\ee
where $\sigma^i = (\mathbbm{1},- i \sigma^a)$ and $\bar{\sigma}^i =
(\mathbbm{1}, i \sigma^a)$ (with $\sigma^a$ the usual Pauli
matrices). Similarly, the hypermultiplet  with physical fields $(\varphi_\alpha, \lambda^{\dot{\alpha}})$ has the  supersymmetry transformations
\bea \label{susytwisted} \delta \varphi_{\alpha} =  \bar{\epsilon}^i
\lambda^{\dot{\alpha}} \sigma^{i}_{\alpha \dot{\alpha}} \,, \qquad
\delta \lambda^{\dot{\alpha}}= \tfrac{1}{4} \gamma^\mu
\partial_\mu \varphi_{\alpha} \epsilon^i \bar{\sigma}^{i\, \dot{\alpha}
\alpha} \,. \eea
These transformations leave invariant the
following action with $8+8$ on-shell degrees of freedom:
\be \label{actionhyptwist} S  =  \int d^3 x \, \left(-
\partial^\mu \varphi^{\dot{\alpha}}
\partial_\mu \varphi_{\dot{\alpha}} - \partial^\mu \varphi_{\alpha}
\partial_\mu \varphi^{\alpha} - 4 \bar{\lambda}^\beta \gamma^\mu
\partial_\mu \lambda_\beta  - 4 \bar{\lambda}_{\dot{\beta}} \gamma^\mu
\partial_\mu \lambda^{\dot{\beta}} \right) \,, \ee
where
\be
\varphi_{\dot{\alpha}} = (\varphi^{\dot{\alpha}})^*\, , \qquad \varphi^\alpha
= (\varphi_\alpha)^*\, , \qquad \bar{\lambda}^\beta = i (\lambda_\beta)^\dag
\gamma^0\, , \qquad \bar{\lambda}_{\dot{\alpha}} = i (\lambda^{\dot{\alpha}})^\dag
\gamma^0\, .
\ee

After coupling to conformal supergravity, the action
\eq{actionhyptwist} gets extended by many terms which are
schematically of the form
\begin{equation}
(\text{matter})^2\times\big\{1+ (\text{Weyl}) +
(\text{Weyl})^2\big\}\,,
\end{equation}
where the terms in brackets denote factors that are independent,
linear or bilinear in the fields of the conformal supergravity
multiplet. In particular, the $(\text{matter})^2\times(\text{Weyl})^2$ terms contain a term proportional to $R \varphi^2$, and this gives rise to the Einstein-Hilbert term on fixing the conformal gauge. Similarly, there will be a term ${\bar\psi}\phi \varphi^2$ from which the gravitino kinetic term follows, after fixing the conformal gauge, on substitution for $\phi_\mu^i$.

It turns out that many terms in the action are irrelevant for the
determination of the final quadratic expression. To keep things simple we only present
those terms in the action that contribute to the final answer. These terms, collectively denoted by
${\cal L}^{{\cal N}=4}_{\text{quadr.}}$, are given by

\begin{eqnarray}\label{actionN=4}
{\cal L}^{{\cal N}=4}_{\text{quadr.}} &=&
-8D^\mu \varphi^{\dot{\alpha}}
D_\mu \varphi_{\dot{\alpha}} -8 D^\mu \varphi_{\alpha}
D_\mu \varphi^{\alpha} - 32 \bar{\lambda}^\beta \gamma^\mu
D_\mu \lambda_\beta  - 32 \bar{\lambda}_{\dot{\beta}} \gamma^\mu
D_\mu \lambda^{\dot{\beta}} \nonumber\\[.2truecm]
&-&  R \varphi_+^2
-\tfrac{1}{2}{\bar\psi}_\mu^k\gamma^{\mu\nu}\phi_\nu^k\,\varphi_+^2
-64  D \varphi_-^2
-32E^2\varphi_+^2\nonumber\\[.2truecm]
&+& 16 {\bar\chi}^i({\bar \sigma}^{i\,\dot \alpha \alpha}
\lambda_\alpha \varphi_{\dot\alpha} - \sigma^i_{\alpha\dot\alpha}\lambda^{\dot\alpha} \varphi^{\alpha})\,,
\end{eqnarray}
where $\varphi_+^2$ and $\varphi_-^2$ are defined in \eqref{defs}.

Note that the $R\varphi^2$ term is precisely such that the scalar wave equation is conformally coupled to the background metric.
Note also that the action is invariant under the exchange of the two $\mathrm{SU}(2)$ factors of the $R$-symmetry group because the $D$, $E$ and $\chi^i$ fields are odd under this exchange. The covariant derivatives $D_\mu$ are covariant with respect to
Lorentz and $\text{SO}(4)$ rotations only. Defining
\begin{equation}V_\mu^{ij} = (\sigma^{ij})_{\alpha}{}^{\beta}V_\mu{}^{\alpha}{}_{\beta}+
({\bar\sigma}^{ij})_{\dot\alpha}{}^{\dot\beta}V_{\mu}{}^{\dot\alpha}{}_{\dot\beta}\;,
\end{equation}
the covariant derivative $D_\mu\phi_\alpha$ is given by

\begin{equation}
D_\mu\varphi_\alpha = \partial_\mu\varphi_\alpha -V_{\mu\,\alpha}{}^{\beta}\varphi_\beta\,.
\end{equation}
In principle, all coefficients in the action \eq{actionN=4} follow from the supersymmetry of the full non-linear action.
In practice, it is much easier to take the action \eq{actionN=4} only as a guideline to obtain the final
quadratic expression. The coefficients in this final expression are easily determined by requiring (linearized) supersymmetry.

After coupling to conformal supergravity the fields of the two hypermultiplets transform under local dilatations and $S$-supersymmetry
transformations as follows:
\begin{eqnarray}
\delta\varphi^{\dot\alpha } &=&\omega \varphi^{\dot \alpha }\,,\hskip 1.5truecm
\delta\lambda_{\alpha} \ = \ \sigma^i_{\alpha\dot\beta}\varphi^{\dot\beta}\eta^i\,,\\[.2truecm]
\delta\varphi_{\alpha } &=& \omega \varphi_{ \alpha }\,,\hskip 1.5truecm
\delta\lambda^{\dot\alpha}  \ = \ {\bar\sigma}^{i\,{\dot\alpha}\beta}\varphi_{\beta}\eta^i\,.
\end{eqnarray}
To fix the dilatations and $S$-supersymmetry transformations we
impose the gauge choices
\begin{equation}\label{Dgauge4}
\varphi_+^2  = 1\,,\hskip 3truecm
{\bar \sigma}^{i\,\dot \alpha \alpha}
\lambda_\alpha \varphi_{\dot\alpha} + \sigma^i_{\alpha\dot\alpha}\lambda^{\dot\alpha} \varphi^{\alpha}=0\,,
\end{equation}
respectively. These gauge choices can be used in  \eq{actionN=4}.
The $D$ and $\chi$
fields are Lagrange multipliers that lead to a single bosonic constraint
\begin{equation}\label{1constraint}
\Phi \equiv \varphi_-^2=0\,
\end{equation}
and 4  fermionic constraints
\begin{equation}\label{4constraints}
\psi^i\equiv {\bar \sigma}^{i\,\dot \alpha \alpha}
\lambda_\alpha \varphi_{\dot\alpha} - \sigma^i_{\alpha\dot\alpha}\lambda^{\dot\alpha} \varphi^{\alpha}=0\,,
\end{equation}
respectively. Since we are going to add further higher-derivative terms to the action in the next subsection
we leave these Lagrange multipliers in the action.

Using the gauge choices \eq{Dgauge4}, the field equation of the
6 vector fields allows for these vector fields to be solved in
terms of the scalars:
\be \label{vecphi} V_\mu^{ij} = (\sigma^{ij})_{\alpha}{}^{ \beta}
\varphi^\alpha
\partial_\mu \varphi_\beta + (\bar{\sigma}^{ij})_{ \dot{\alpha}}{}^{
\dot{\beta}} \varphi^{\dot{\alpha}}\partial_\mu \varphi_{\dot{\beta}} \,.
\ee
We next fix the local $\text{SO}(4)$ by imposing the following 6
conditions:
\begin{equation}\label{SO4gauge}
\varphi^\alpha\partial_\mu\varphi_\beta + \text{h.c.} =0\,,\hskip 2truecm
\varphi^{\dot\alpha}\partial_\mu\varphi_{\dot\beta}+\text{h.c.}=0\,,
\end{equation}
which implies $V_\mu^{ij}=0$.

We note that for pure Einstein supergravity (no further higher-derivative terms added to the action) the 8+8 matter
fields of the compensating hypermultiplets are all fixed by the constraints and gauge choices,
as the following counting shows:

\begin{eqnarray}
&&\text{bosons}\,:\ \ 8 - 1\ (\text{D-gauge})  - 6\ (\text{SO(4)-gauge })- 1\ (\text{constraint}) = 0\,,
\nonumber\\[.2truecm]
\hskip 1truecm
&&\text{fermions}\,:\ \ 8 - 4 \ (\text{S-gauge}) - 4 \ (\text{constraints}) = 0\,.
\end{eqnarray}

Using the gauge choices \eq{Dgauge4} and \eq{SO4gauge}, one can show that it
is possible to rewrite the kinetic terms of the compensating matter scalars and
fermions, appearing in the conformal action, as  standard kinetic
terms for $\Phi$ and $\psi^i$ only. One thus obtains the following
linerarized ${\cal N}=4$ supersymmetric Einstein action

\begin{eqnarray} \label{n4eh}
S^{{\cal N}=4}_{2} &=& \int\,d^3 x\,\bigg\{\frac{1}{2} h^{\mu \nu}
G^{(\mathrm{lin})}_{\mu \nu} +
\epsilon^{\mu\nu\rho}{\bar\psi}_\mu^i\partial_\nu\psi_\rho^i -
V_\mu^{ij}V_{\mu\,ij}
-32 E^2 \nonumber\\[.2truecm]
&& \hskip 2truecm -8 {\bar\psi}\partial \hskip -.24truecm  /\,\psi
+16   {\bar\psi}\chi -64 D\Phi +32 \Phi\Box \Phi\bigg\}\,.
\end{eqnarray}

Using the fact that the gauge-fixing requires a compensating
$S$-transformation and $\text{SO}(4)$ transformation with parameters
\begin{equation}
\eta^i= \frac{1}{2}\gamma^\rho V_\rho^{ij}\epsilon^j  \,,\hskip
2truecm \Lambda^{ij}= -{\bar\epsilon}^{[i}\psi^{j]}   \,,
\end{equation}
respectively, we find that the action \eq{n4eh} is invariant
under the following supersymmetry rules
\begin{eqnarray}\label{N=4lintr}
\delta h_{\mu\nu} &=& {\bar\epsilon}^i\gamma_{(\mu}\psi_{\nu)}^i\,,\nonumber\\[.2truecm]
\delta\psi_\mu^i &=& -\tfrac{1}{4}\gamma^{\rho\sigma}\partial_\rho
h_{\mu\sigma}\epsilon^i
- V_\mu^{ij}\epsilon^j + \tfrac{1}{2} \gamma_\mu \gamma^\rho V_{\rho}^{ij} \epsilon^{j}\,,\nonumber  \\[.2truecm]
\delta V_\mu^{ij} &=& \frac{1}{2} {\bar\epsilon}^{[i}\phi_\mu^{j]} +
\epsilon^{ijkl} {\bar\epsilon}^k \gamma_\mu\chi^l - \epsilon^{ijkl}
\bar{\epsilon}^k \partial_\mu \psi^l \,, \hskip 1truecm
\delta E =  \frac{1}{4} {\bar\epsilon}^{i}\chi^{i}\,,   \\[.2truecm]
\delta\chi^{i} &=& \frac{1}{8} \epsilon^{ijkl}\gamma^\mu
F_{\mu\,\text{(lin)}}^{jk}\epsilon^{l} +  \gamma^\mu(\partial_\mu
E)\epsilon^i + D\epsilon^i\,, \hskip .5truecm \delta D =
\frac{1}{4}{\bar\epsilon}^{i} \partial \hskip -.24truecm  /
\,\chi^{i}\,,               \nonumber\\ [.2truecm] \delta\psi^{i}
&=& \frac{1}{8}\epsilon^{ijkl} \gamma^\mu V_\mu^{jk}\epsilon^{l} +
E\epsilon^i + \gamma^\mu(\partial_\mu \Phi)\epsilon^i\,,\hskip
1truecm \delta \Phi =  \frac{1}{4}{\bar
\epsilon}^{i}\psi^{i}\,.\nonumber
\end{eqnarray}

Note that the action \eq{n4eh} does not describe any massive
degrees of freedom. Several fields could
be integrated out to leave us with only the Einstein term and
the kinetic term of the gravitino. We do not do this because the equations of
motion of these fields change as soon as we add one of the
superconformal higher-derivative actions constructed in the previous subsection.


\subsection{${\cal N}=4$ massive supergravities}

In this subsection we will consider the sum of the supersymmetric Einstein action and
the conformal higher-derivative actions constructed in subsection 3.2 with 3,4
and 5 derivatives thereby introducing three mass parameters $\mu\,,m$ and $M$, respectively.
This leads to the following action that is left invariant under the same transformation rules
\eqref{N=4lintr} derived in the previous subsection:
\begin{eqnarray}
S^{{\cal N}=4} &=& \int\,d^3 x\,\bigg\{\frac{1}{2} h^{\mu \nu}
G^{(\mathrm{lin})}_{\mu \nu} +
\epsilon^{\mu\nu\rho}{\bar\psi}_\mu^i\partial_\nu\psi_\rho^i -
V_\mu^{ij}V_{\mu\,ij}
-32 E^2 \nonumber\\[.2truecm]
&& \hskip 2truecm - 8 {\bar\psi}\partial \hskip -.24truecm  /\,\psi
+ 16   {\bar\psi}\chi
- 64 D\Phi + 32 \Phi\Box \Phi\bigg\} \nonumber\\[.2truecm]
&&+ \frac{1}{\mu} \bigg\{h^{\mu\nu}C_{\mu\nu}^{\rm (lin)} +
{\bar\psi}^i_\mu{\cal C}_{\rm (lin)}^{\mu\,i}
-2 V_\mu^{ij}F_{\rm (lin)}^{\mu\,ij} +16{\bar\chi}^{i}\chi^{i}-128 ED\bigg\}\nonumber\\[.2truecm]
&&+\frac{1}{m^2}\bigg\{-\tfrac{1}{2}\epsilon^{\mu\tau\rho}h_\mu{}^\nu\partial_\tau
C_{\rho\nu}^{\rm (lin)} -\frac{1}{2} {\bar\psi}^i_\mu \partial
\hskip -.24truecm  / {\cal C}^{{i\, \mu\, \rm (lin)}}
+  F_{\mu\,{\rm (lin)}}^{ij} F_{\rm (lin)}^{\mu\,ij} +\\[.2truecm]
&&\hskip 3truecm +32 E\Box E -8 {\bar\chi}^{i}\partial \hskip -.24truecm  /\,\chi^{i}+32 D^2\bigg\}\nonumber\\[.2truecm]
&&+\frac{1}{M^3} \bigg\{R^{\mu\nu}_{\text{(lin)}} C_{\mu\nu}^{\rm (lin)} + {\bar {\cal C}}_{\mu\, i}^{\text{(lin)}} {\cal C}_{\rm (lin)}^{\mu\,i}+\epsilon^{\mu\nu\rho}F_\mu^{\text{(lin)}\, ij}\partial_\nu F^{\text{(lin)}\, ij}_{\rho} \nonumber\\[.2truecm]
&&\hskip 3truecm +16{\bar\chi}^{i}\Box \chi^{i}-128 E\Box
D\bigg\}\,.\nonumber
\end{eqnarray}

To analyze which massive supermultiplets are propagating for the
different values of the mass parameters, it is enough to consider
the scalar equations of motion:
\begin{eqnarray}
E+\frac{2}{\mu}D-\frac{1}{m^2}\Box E+\frac{2}{M^3}\Box D &=&0\,,\nonumber\\[.2truecm]
\Phi+\frac{2}{\mu}E-\frac{1}{m^2}D+\frac{2}{M^3}\Box E&=&0\,,\\[.2truecm]
\Box\Phi - D&=&0\,,\nonumber
\end{eqnarray}
and the fermionic equations of motion:
\begin{eqnarray}
& & \partial \hskip -.24truecm  / \psi^i  =  \chi^i \,, \nonumber \\
& & \psi^i + \frac{2}{\mu} \chi^i - \frac{1}{m^2} \partial \hskip
-.24truecm  / \chi^i + \frac{2}{M^3} \Box \chi^i = 0 \,.
\end{eqnarray}

We may now analyze the following models
\bigskip
\bigskip

\noindent (1)\ \ {\sl Supersymmetric Einstein}\,: $\mu\,, m\,,
M\rightarrow \infty$
\bigskip

\noindent We find
\begin{equation}
E=D=\Phi=\psi^i=\chi^i=0\;,
\end{equation}
and hence there are no propagating multiplets.
\bigskip
\bigskip

\noindent (2)\ \ {\sl Supersymmetric TMG}\,: $ m\,, M\rightarrow
\infty$
\bigskip

\noindent We find one independent propagating scalar that satisfies
\begin{equation}
\Box\Phi - \frac{\mu^2}{4}\Phi=0\;,
\end{equation}
and hence there is one propagating massive spin 2 supermultiplet. For the
fermions we find
\begin{equation}
\dslash \psi^i = \chi^i \,, \qquad \psi^i = - \frac{2}{\mu} \chi^i
\,,
\end{equation}
and hence
\begin{equation}
\dslash \psi^i = - \frac{\mu}{2} \psi^i \,.
\end{equation}
Applying an extra $\dslash$ leads to
\begin{equation}
\Box \psi^i - \frac{\mu^2}{4}\psi = 0 \,,
\end{equation}
which is in agreement with the ${\cal N}=4$ spin 2 supermultiplet content displayed in Table~1.

\bigskip
\bigskip

\noindent (3)\ \ {\sl Supersymmetric NMG}\,: $ \mu\,, M\rightarrow
\infty$
\bigskip

\noindent We find two independent scalars with
\begin{equation}
\Box D - m^2 D = 0\,,\hskip 2truecm \Box E - m^2 E=0\,,
\end{equation}
and hence there are two propagating ${\cal N}=4$  massive multiplets, of the same
mass but with opposite helicity. For the fermions we find: 
\begin{equation}
\dslash \psi^i = \chi^i \,, \qquad \dslash \chi^i = m^2 \psi^i \,.
\end{equation}
%
Upon diagonalization of these equations one infers that their mass eigenvalues are $\pm m$, which
implies that they describe opposite helicities, as required.
By applying $\dslash$, one finds
\begin{equation}
\Box \chi^i - m^2 \chi^i = 0 \,, \qquad \Box \psi^i - m^2 \psi^i = 0
\,,
\end{equation}
which is again in agreement with the scalar analysis.

\bigskip
\bigskip

\noindent (4)\ \ {\sl Supersymmetric GMG}\,: $ M\rightarrow \infty$
\bigskip

\noindent We now have
\begin{eqnarray}
E+\frac{2}{\mu}D-\frac{1}{m^2}\Box E&=&0\,,\nonumber\\[.2truecm]
\Phi+\frac{2}{\mu}E-\frac{1}{m^2}D&=&0\,,\\[.2truecm]
\Box\Phi - D&=&0\,.\nonumber
\end{eqnarray}
Eliminating the scalar $\Phi$ and replacing $E\rightarrow 1/\mu E$
we obtain
\begin{eqnarray}
\Box D - (m^2+\frac{4 m^4}{\mu^2})D - \frac{2 m^4}{\mu^2} E&=&0\,,\nonumber\\[.2truecm]
\Box E - 2 m^2 D - m^2 E&=&0\,.
\end{eqnarray}
After a diagonalization we find  two propagating multiplets with
different masses :
\begin{equation}
m_\pm^2 = \frac{m^2 \Big(2m^2 + \mu^2 \pm 2 \sqrt{m^2(m^2 +
\mu^2)}\Big)}{\mu^2}\;,
\end{equation}
such that\footnote{We have fixed the sign of $m_--m_+$ in accordance
with \cite{Bergshoeff:2009hq}. To compare, use that $\tilde\mu = 2m^2/\mu^2$.}
\begin{equation}
m_+m_- = m^2\,,\hskip 2truecm m_--m_+ = 2\frac{m^2}{\mu}\,.
\end{equation}
The corresponding eigenscalars are $D + b_\pm E$, where
\begin{equation}
b_\pm = \frac{-m^2 \pm \sqrt{m^4 + m^2 \mu^2}}{\mu^2}\,.
\end{equation}

For the fermions we find
\begin{eqnarray}
& & \dslash \psi^i = \chi^i \,, \nonumber \\
& & \dslash \chi^i = m^2 \psi^i + \frac{2 m^2}{\mu} \chi^i \,.
\end{eqnarray}
By applying $\dslash$ and similar manipulations as done for the
scalars, one arrives at
\begin{eqnarray}
& & \Box \psi^i - 2 m^2 \chi^i - m^2 \psi^i = 0 \,, \nonumber \\
& & \Box \chi^i - \Big(m^2 + \frac{4 m^4}{\mu^2}\Big) \chi^i -
\frac{2 m^4}{\mu^2} \psi^i = 0 \,.
\end{eqnarray}
This is the same system of equations as encountered in the scalar
case, with $D \rightarrow \chi^i$, $E \rightarrow \psi^i$. The
diagonalisation procedure and mass$^2$ eigenvalues are thus the same
as in the scalar case.

\bigskip
\bigskip

\noindent (5)\ \ {\sl Supersymmetric NTMG}\,: $ M\rightarrow
\infty\,, \mu\,,m\rightarrow 0$\ such that $m^2/\mu=\text{const.}$
\bigskip

\noindent We now have one propagating scalar,
\begin{equation}
\Box E - \frac{4 m^4}{\mu^2}E = 0\;,
\end{equation}
and hence one propagating multiplet. Similarly, we find for the
fermions:
\begin{equation}
\dslash \psi^i = \chi^i\,, \qquad \dslash \chi^i = \frac{2 m^2}{\mu}
\chi^i \,,
\end{equation}
from which one derives
\begin{equation}
\Box \chi^i - \frac{4 m^4}{\mu^2} \chi^i = 0 \,,
\end{equation}
again in agreement with the scalars.

\bigskip
\bigskip

\noindent (6)\ \ {\sl Supersymmetric ENMG}\,: $ m^2\rightarrow
\infty\,, \mu\,,M\rightarrow 0$\ such that $\mu/M^3=\text{const.}$
\bigskip

\noindent This is the case of `Extended New Massive Gravity' (ENMG) discussed in the third reference of
\cite{Bergshoeff:2009hq}. This case leads to the scalar equations
\begin{equation}
\frac{1}{\mu}D +\frac{1}{M^3}\Box D=0\,,\hskip 2truecm
\frac{1}{\mu}E +\frac{1}{M^3}\Box E=0\,.
\end{equation}
Since in the action we have
\begin{equation}
E\Box D \sim A\Box A - B\Box B\,,\hskip 2truecm E=A+B\,, D=A-B\,,
\end{equation}
we end up with one physical and one ghost multiplet, each with
$\text{(mass)}^2 = M^3/\mu$.

For the fermions we find
\begin{equation}
 \frac{1}{\mu} \chi^i + \frac{1}{M^3} \Box \chi^i = 0 \,,
\end{equation}
which also implies one physical and one ghost multiplet.
\bigskip

We have summarized the analysis of the different models in Table 2.

\bigskip

\begin{table}\label{multipletsn4}
\caption{{\footnotesize{Some 3D higher-derivative sugra models. The
second column indicates the number of derivatives in the different
terms for the spin 2 field. The 3rd and 4th column indicate the
massive spin 2 multiplets described by the corresponding
supergravity model. The last 2 columns indicate the masses of these
${\cal N}=4$ multiplets, respectively. The boldface number in the
lowest row indicates a ghost multiplet.}}} \centering
\begin{tabular}{cccccccccc}
\\[0.02cm]
\hline\hline\\[-.2truecm]
sugra model&action&${\cal N}=4$&opposite helicity&(mass)${}^2$&(mass)${}^2$\\[.2truecm]
\hline\\[-.2truecm]
Einstein& 2&--&--&--&--\\[.2truecm]
TMG& 2+3&1&--&$\mu^2$&--\\[.2truecm]
NMG&2+4&1&1&$m^2$&$m^2$\\[.2truecm]
GMG& 2+3+4&1&1&$m_+^2$&$m_-^2$\\[.2truecm]
NTMG& 3+4&1&--&$m^4/\mu^2$&--\\[.2truecm]
ENMG& 3+5&1&{\bf 1}&$M^3/\mu$&$M^3/\mu$\\[.2truecm]\hline
\end{tabular}
\end{table}
\bigskip


\section{$\mathcal{N}=8$ Maximal Supergravity}

In this section we consider the case of maximal supersymmetry. Since the application of the superconformal tensor calculus
has been explained in the previous section we will be brief whenever there is overlap with the ${\cal N}=4$ case.


\subsection{The $\mathcal{N}=8$ conformal multiplet}

The $D=3$, ${\cal N}=8$ linearized conformal supergravity multiplet
has 128+128 field degrees of freedom and contains the following
components:
\begin{equation}\label{fields}
\big \{h_{\mu\nu}\,, \psi_\mu^i\,, V_\mu^{ij}\,, E^{ijkl}\,,
\chi^{ijk}\,, D^{ijkl}\big \}\,,
\end{equation}
with $h_{\mu\nu}=h_{\nu\mu}$ the linearized graviton, $\psi_\mu^i\, (i=1,\cdots , 8)$ the
8 gravitini and
$V_\mu^{ij}=-V_\mu^{ji}$ the ${\rm SO}(8)$
R-symmetry gauge field. The matter fields $E^{ijkl}$ and $D^{ijkl}$
(of different mass dimension) are in the antisymmetric selfdual
${\bf 35^+}$ and anti-selfdual ${\bf 35^-}$ representation of ${\rm
SO}(8)$, respectively, while the fermions $\chi^{ijk}$ form an
antisymmetric ${\bf 56}$\,-plet of ${\rm SO}(8)$.

The linearized supersymmetry transformations are given by \cite{Howe:1995zm} (using the
same conventions as in \cite{Andringa:2009yc})
\begin{eqnarray}\label{lintrrules}
\delta h_{\mu\nu} &=& {\bar\epsilon}^i\gamma_{(\mu}\psi_{\nu)}^i\,,\nonumber\\[.2truecm]
\delta\psi_\mu^i &=& -\tfrac{1}{4}\gamma^{\rho\sigma}\partial_\rho
h_{\mu\sigma}\epsilon^i
- V_\mu^{ij}\epsilon^j \,,\nonumber  \\[.2truecm]
\delta V_\mu^{ij} &=& \tfrac{1}{2} {\bar\epsilon}^{[i}\phi_\mu^{j]}   +  {\bar\epsilon}^k \gamma_\mu\chi^{ijk} \,, \nonumber \\[.2truecm]
\delta E^{ijkl} &=&  - {\bar\epsilon}^{[i}\chi^{jkl]}\ +\ \text{dual}\,,   \\[.2truecm]
\delta\chi^{ijk} &=& -\tfrac{3}{4} \gamma^\mu F_{\mu\,\text{(lin)}}^{[ij}\epsilon^{k]} +  \gamma^\mu(\partial_\mu E^{ijkl})\epsilon^l  + D^{ijkl}\epsilon^l\,, \nonumber \\[.2truecm]
\delta D^{ijkl} &=& -{\bar\epsilon}^{[i} \partial \hskip -.24truecm
/ \,\chi^{jkl]} \ -\ \text{dual}\,.               \nonumber
\end{eqnarray}
Here $F^{\mu\,ij}_{\rm (lin)}$ and $\phi_\mu^i$ are defined in an
analogous manner as in \eq{defFphi} for $\mathcal{N}=4$. We  insist
on gauge invariance with respect to the analogue of the linear gauge
transformations of \eq{lineargauge1}. Finally, the supersymmetry
algebra  closes in the same way as in the ${\cal N}=4$ case.


\subsection{$\mathcal{N}=8$ conformal higher-derivative actions}

Using the same definitions of the Cotton and Cottino tensor as in the ${\cal N}=4$ case we may construct the following parity-even
conformal higher-derivative actions for the
${\cal N}=8$ conformal multiplet.
\bigskip

\noindent (1) ${\cal N}=8$ {\sl supersymmetric} $R^2$
\vskip .2truecm

There exists a supersymmetric action of the ${\cal N}=8$ conformal
multiplet that starts with the linearized version of the
$R_{\mu\nu}^2-\tfrac{3}{8}R^2$ term of NMG
\cite{Bergshoeff:2009hq}.  More precisely, the
transformation rules \eqref{lintrrules} leave the following action
invariant:
\begin{eqnarray}\label{Kaction}
S_4^{{\cal N}=8} &=& \int
d^3x\,\bigg\{-\frac{1}{2}\epsilon^{\mu\tau\rho}h_\mu{}^\nu\partial_\tau
C_{\rho\nu}^{\rm (lin)} -\frac{1}{2} {\bar\psi}^i_\mu  \partial
\hskip -.24truecm  / {\cal C}^{{\mu\,\rm (lin)\,i}}
+  F_{\mu\,{\rm (lin)}}^{ij} F_{\rm (lin)}^{\mu\,ij} +\nonumber\\[.2truecm]
&&+ \frac{2}{3} E^{ijkl}\Box E_{ijkl}  - \frac{4}{3}
{\bar\chi}^{ijk}\partial \hskip -.24truecm  /\,\chi^{ijk}+
\frac{2}{3} D^{ijkl} D^{ijkl}\bigg\}\,.
\end{eqnarray}

Under the rigid
supersymmetry rules \eqref{lintrrules} the Cotton, Cottino and $\text{SO(8)}$ curvature tensors  transform as follows
\begin{eqnarray}
\delta C_{\mu\nu}^{\rm (lin)} &=&  -\tfrac{1}{4}{\bar\epsilon}^i\gamma_{(\mu}{}^\rho\partial_\rho{\cal C}_{\nu)}^{i\,{\rm lin}}\,,   \nonumber\\[.2truecm]
\delta{\cal C}^{\mu\, i}_{\rm (lin)} &=&  \gamma_\nu\epsilon^i
C^{\mu\nu}_{\rm (lin)}
+\epsilon^{\mu\nu\rho}\gamma_\sigma\gamma_\nu\epsilon^j\partial_\rho F^{\sigma\,ij}_{\rm (lin)}\,,  \\[.2truecm]
\delta F^{\mu\,ij}_{\rm (lin)} &=&
\tfrac{1}{2}{\bar\epsilon}^{[i}{\cal C}^{\mu\, j]}_{\rm (lin)} +
{\bar\epsilon}^k\gamma^{\mu\rho}\partial_\rho\chi^{ijk}\,.
\nonumber
\end{eqnarray}
These transformation rules define a ${\cal N}=8$ conjugated Weyl multiplet with
components
\begin{equation}\label{Weylc}
\big \{C_{\mu\nu}^{\rm (lin)}\,,{\cal C}^{\mu\, i}_{\rm
(lin)}\,,F^{\mu\,ij}_{\rm (lin)}\,,\chi^{ijk}\,,D^{ijkl}\,, E^{ijkl}
\big\}\,.
\end{equation}

\bigskip

\noindent (2) $\mathcal{N}=8$ {\sl supersymmetric Cotton tensor squared }
\vskip .2truecm

Following the same procedure as in the ${\cal N}=4$ case, making use of the conjugated Weyl multiplet \eq{Weylc},
we obtain the following supersymmetric Cotton tensor squared action:
\begin{eqnarray}\label{Caction}
S_6^{{\cal N}=8} &=& \int d^3x\,\bigg\{C_{\rm
(lin)}^{\mu\nu}C_{\mu\nu}^{\rm (lin)} - \frac{1}{4} {\bar {\cal
C}}^{i\,{\rm (lin)}}_\mu  \partial \hskip -.24truecm  / {\cal
C}_{\rm (lin)}^{\mu\,i}
+ F_{\mu\,{\rm (lin)}}^{ij}\Box F_{\rm (lin)}^{\mu\,ij} +\nonumber\\[.2truecm]
&& +\frac{2}{3} E^{ijkl}\Box^2E^{ijkl} -
\frac{4}{3}{\bar\chi}^{ijk}\Box\,\partial \hskip -.24truecm
/\,\chi^{ijk}+\frac{2}{3} D^{ijkl}\Box D^{ijkl} \bigg\}\,.
\end{eqnarray}
\bigskip

This concludes our discussion of the (parity-even) conformal higher-derivative actions.
In the next subsection we consider the case of (non-conformal) ${\cal N}=8$ Einstein supergravity.


\subsection{$\mathcal{N}=8$ Einstein supergravity}

To obtain supersymmetric Einstein we introduce $8$ scalar multiplets
and couple them to conformal supergravity. The supersymmetry
transformation rules of $8$ scalar multiplets with components
$(\phi^{\alpha A},\chi^A_{\dot\alpha})\ (A=1,\cdots, 8)$ are given
by
\begin{eqnarray}\label{susyscalar}
\delta\varphi^{\alpha A} &=& {\bar\epsilon}^i(\gamma^i)^{\alpha{\dot\beta}}\lambda^A_{\dot\beta }\,,\nonumber\\[.2truecm]
\delta\lambda^A_{\dot\alpha }
&=&\frac{1}{4}\gamma^\mu(\partial_\mu\varphi^{\alpha A}){\tilde
\gamma}^i_{{\dot\alpha}\alpha}\epsilon^i\,,
\end{eqnarray}
where $(i,\alpha,{\dot\alpha})= (1,\cdots,8)$ denote the
8-dimensional ${\bf v}, {\bf s}$ and ${\bf c}$ representations of
$\text{SO}(8)$, respectively. For the $\text{SO}(8)$ Dirac matrices
we use the notation of \cite{Green:1982tc}.

The transformation rules \eq{susyscalar} leave the following action
invariant
\begin{equation}
S = \int d^3x\,\big\{\partial_\mu\varphi^{\alpha
A}\partial^\mu\varphi^A_\alpha + 4{\bar\lambda}^{\dot\alpha A}\partial
\hskip -.24truecm  / \lambda^A_{\dot\alpha}\big\}\,.
\end{equation}
We next couple the above action to ${\cal N}=8$ conformal supergravity and
follow precisely the same steps as in the ${\cal N}=4$ case. From now on we
 identify $A=\alpha$ and write $\varphi^{\alpha B} = \varphi^{\alpha\beta}$. This leads to
the following Einstein supergravity action
\begin{eqnarray}\label{action4}
S^{{\cal N}=8}_{2} &=& \int\,d^3 x\,\bigg\{\frac{1}{2} h^{\mu \nu}
G^{(\mathrm{lin})}_{\mu \nu} +
\epsilon^{\mu\nu\rho}{\bar\psi}_\mu^i\partial_\nu\psi_\rho^i -
V_\mu^{ij}V_{\mu\,ij}
-\frac{2}{3} E^{ijkl}E_{ijkl} \nonumber\\[.2truecm]
&&  - \frac{4}{3} {\bar\psi}^{ijk}\partial \hskip -.24truecm
/\,\psi^{ijk}  +\frac{8}{3}   {\bar\psi}^{ijk}\chi_{ijk}
-\frac{4}{3} D^{ijkl}\Phi_{ijkl} +\frac{2}{3} \Phi^{ijkl}\Box
\Phi_{ijkl}\bigg\}\,.
\end{eqnarray}
We find that this action is invariant under the following supersymmetry rules:
\begin{eqnarray}\label{N=8NMG2dversion}
\delta h_{\mu\nu} &=& {\bar\epsilon}^i\gamma_{(\mu}\psi_{\nu)}^i\,,\nonumber\\[.2truecm]
\delta\psi_\mu^i &=& -\tfrac{1}{4}\gamma^{\rho\sigma}\partial_\rho
h_{\mu\sigma}\epsilon^i
- V_\mu^{ij}\epsilon^j + \tfrac{1}{2} \gamma_\mu \gamma^\rho V_{\rho}^{ij} \epsilon^{j}\,,\nonumber  \\[.2truecm]
\delta V_\mu^{ij} &=& \tfrac{1}{2} {\bar\epsilon}^{[i}\phi_\mu^{j]}
+  {\bar\epsilon}^k \gamma_\mu\chi^{ijk}
- \bar{\epsilon}^k \partial_\mu \psi^{ijk} \,, \nonumber \\[.2truecm]
\delta E^{ijkl} &=&  - {\bar\epsilon}^{[i}\chi^{jkl]}\ +\ \text{dual}\,,   \\[.2truecm]
\delta\chi^{ijk} &=& -\tfrac{3}{4} \gamma^\mu F_\mu^{[ij}\epsilon^{k]} +  \gamma^\mu(\partial_\mu E^{ijkl})\epsilon^l  + D^{ijkl}\epsilon^l\,, \nonumber \\[.2truecm]
\delta D^{ijkl} &=& -{\bar\epsilon}^{[i} \partial \hskip -.24truecm
/ \,\chi^{jkl]} \ -\ \text{dual}\,,               \nonumber\\
[.2truecm] \delta\psi^{ijk} &=& -\tfrac{3}{4} \gamma^\mu
V_\mu^{[ij}\epsilon^{k]} +   E^{ijkl}\epsilon^l +
\gamma^\mu(\partial_\mu \Phi^{ijkl})\epsilon^l\,,\nonumber\\
[.2truecm] \delta \Phi^{ijkl} &=& -{\bar \epsilon}^{[i}\psi^{jkl]} \
-\ \text{dual}\,.\nonumber
\end{eqnarray}


\subsection{$\mathcal{N}=8$ new massive supergravity}

To obtain $\mathcal{N}=8$ new massive supergravity we add the
supersymmetric $R^2$ action \eq{Kaction}, which we give a
coefficient $1/m^2$,  to the supersymmetric Einstein action. Since the conformal multiplet is off-shell
we retain supersymmetry. The combined action is given by
\begin{eqnarray}
S^{{\cal N}=8}_{\text{NMG}} &=& \int\,d^3 x\,\bigg\{\frac{1}{2}
h^{\mu \nu} G^{(\mathrm{lin})}_{\mu \nu} +
\epsilon^{\mu\nu\rho}{\bar\psi}_\mu^i\partial_\nu\psi_\rho^i -
V_\mu^{ij}V_{\mu\,ij}
-\frac{2}{3} E^{ijkl}E_{ijkl} \nonumber\\[.2truecm]
&&  - \frac{4}{3} {\bar\psi}^{ijk}\partial \hskip -.24truecm
/\,\psi^{ijk}  +\frac{8}{3}   {\bar\psi}^{ijk}\chi_{ijk}
-\frac{4}{3} D^{ijkl}\Phi_{ijkl} +\frac{2}{3} \Phi^{ijkl}\Box \Phi_{ijkl}\bigg\}\nonumber\\[.2truecm]
&&+\frac{1}{m^2}
\bigg\{-\tfrac{1}{2}\epsilon^{\mu\tau\rho}h_\mu{}^\nu\partial_\tau
C_{\rho\nu}^{\rm (lin)} -\frac{1}{2} {\bar\psi}^i_\mu  \partial
\hskip -.24truecm  / {\cal C}^{{\mu\,\rm (lin)\,i}}
+  F_{\mu\,{\rm (lin)}}^{ij} F_{\rm (lin)}^{\mu\,ij} \nonumber\\[.2truecm]
&& - \frac{4}{3} {\bar\chi}^{ijk}\partial \hskip -.24truecm
/\,\chi^{ijk}
 + \frac{2}{3} E^{ijkl}\Box E_{ijkl} + \frac{2}{3} D^{ijkl}D_{ijkl}
\bigg\}\,.
\end{eqnarray}
It is invariant under the same supersymmetry transformations
\eq{N=8NMG2dversion}. The only thing that changes with respect to the pure Einstein supergravity action
is that
 the
equations of motion of the conformal fields  $E,D,V$ and $\chi$,
receive  $1/m^2$ corrections which leads to propagating massive
degrees of freedom. To be precise, the corrected equations of motion read:
\begin{eqnarray}
E_{ijkl} - \frac{1}{m^2}\Box E_{ijkl} &=& 0\,,\\[.2truecm]
V_{\mu\,ij} - \frac{1}{m^2}\partial^\lambda
F_{\lambda\mu\,ij}^{\text{(lin)}}(V)
 &=& 0\,,\\[.2truecm]
\Phi^{ijkl} - \frac{1}{m^2}D^{ijkl}  &=& 0\,,\\[.2truecm]
\psi^{ijk} - \frac{1}{m^2}\partial \hskip -.24truecm
/\,\chi^{ijk}&=& 0\,.\label{psi}
\end{eqnarray}

The first  equation shows that the $E$ scalars describe 35 massive
helicity 0 d.o.f. The third equation, together with the uncorrected
equation for $\Phi^{ijkl}$,
\begin{equation}
\Box \Phi^{ijkl} - D^{ijkl}=0\,,
\end{equation}
can be used to show that the $D$ scalars satisfy
\begin{equation}
D^{ijkl} - \frac{1}{m^2}\Box D^{ijkl} = 0\,,
\end{equation}
and hence describe another 35 helicity d.o.f. of the same mass $m$.
 From the second equation it follows that the vector fields describe
28 helicity $\pm 1$ states:
\begin{equation}
V_\mu^{ij}-\frac{1}{m^2}\Box V_\mu^{ij}=0\,.
\end{equation}

Concerning  the helicity $\pm 1/2$ degrees of freedom we end up with
the following two equations of motion:
\begin{equation}
\partial \hskip -.24truecm / \chi^{ijk} - m^2\psi^{ijk}=0\,,\hskip 2truecm \partial \hskip -.24truecm / \psi^{ijk} - \chi^{ijk}=0\,.
\end{equation}
By taking sums and differences we see that this system describes 56
$+1/2$ helicity states and 56 $-1/2$ helicity states of the same
mass $m$, i.e.
\begin{equation}
\chi^{ijk}-\frac{1}{m^2}\Box\chi^{ijk}=0\,,\hskip 2truecm
\psi^{ijk}-\frac{1}{m^2}\Box\psi^{ijk}=0\,.
\end{equation}

Adding up all massive degrees of freedom, including the helicity
$\pm 2$ and $\pm 3/2$ states,  we precisely obtain the content of
the ${\cal N}=8$ massive super multiplet given in Table~1, as it
should be.


\section{Linearized $\mathcal{N}=7$ TMG}

In this section we study maximal supersymmetry for parity-odd actions.
We first consider the Lorentz Chern-Simons (LCS) term.
It turns out that there does not exist a  ${\cal N}=8$
supersymmetric LCS action for the {\sl full} conformal
multiplet. The reason for this is that such an action would require
a term of the form $E^{ijkl}D^{ijkl}$ but, due to the opposite
dualities of the $E$ and $D$ fields, such a term does not exist. The
best one can do is write down the following so-called
``pseudo-action'' assuming that $E^{ijkl}$ and $D^{ijkl}$ are not
(anti-)selfdual (the Cotton tensor $C$ and the Cottino tenor ${\cal
C}$ are defined below):
\begin{equation}\label{linaction}
S_3^{{\cal N}=8} = \int d^3x\,\bigg\{h^{\mu\nu}C_{\mu\nu}^{\rm
(lin)} + {\bar\psi}^i_\mu{\cal C}_{\rm (lin)}^{\mu\,i} -2
V_\mu^{ij}F_{\rm (lin)}^{\mu\,ij} + \frac{8}{3}
{\bar\chi}^{ijk}\chi^{ijk}\bigg\}\,.
\end{equation}
This action is invariant under supersymmetry up to terms
proportional to $E^{ijkl}$ or $D^{ijkl}$. Indeed, up to a total
derivative, the variation of the above action is given by
\begin{equation}\label{varlinaction}
\delta S_3^{{\cal N}=8} = \int d^3x\,\bigg\{ \frac{16}{3} {\bar
\chi}^{ijk} \gamma^\mu \epsilon^l \partial_\mu E^{ijkl} +
\frac{16}{3} \bar{\chi}^{ijk} \epsilon^l D^{ijkl}  \bigg\}\,.
\end{equation}
The (anti-)selfduality of the $E$ and $D$ fields is only imposed at
the level of the equations of motion. Note that setting
$E^{ijkl}=D^{ijkl}=0$ in the conformal multiplet leads to equations
of motion for the remaining fields  that can be integrated to an
action which is precisely the above action \cite{Gran:2008qx}. We
have not been able to write down a similar pseudo-action for the
combined Einstein-Chern-Simons system. This is only possible if one
can impose (anti-) selfduality in the equations of motion for
$D^{ijkl}$ and $E^{ijkl}$. This seems unlikely since we expect an
equation of motion of the form $E^{ijkl}=\tfrac{1}{\mu} D^{ijkl} +
\cdots$ where $\mu$ is a mass parameter. Such an equation obviously
is inconsistent with the duality properties of the $E$ and $D$
fields. The non-existence of such a  ${\cal N}=8$  topologically
massive supergravity theory can also be anticipated from the fact
that such a theory would lead to a ${\cal N}=8$ supersymmetric
massive supermultiplet with broken parity. According to Table 1 such
a multiplet  does not exist.

Although a supersymmetric ${\cal N} = 8$ LCS action does
not exist, one can construct a supersymmetric ${\cal N} = 7$
version. In order to do this, one decomposes the $R$-symmetry index
$i$ as $i = \{I, 8\}$, where $I= 1,\cdots, 7$. Performing this
decomposition, one arrives at the following fields:
\begin{eqnarray}
& & h_{\mu \nu} \,, \quad \psi^I_\mu \,, \quad \psi_\mu \equiv \psi^8_\mu \,, \quad V_\mu^{IJ}\, \quad V_\mu^I \equiv V_\mu^{I 8} \,, \quad \chi^{IJK}\,, \quad \chi^{IJ} \equiv \chi^{IJ8} \,, \nonumber \\
& & E^{IJKL} \,, \quad E^{IJK} \equiv E^{IJK8} \,, \quad D^{IJKL}
\,, \quad D^{IJK} \equiv D^{IJK8} \,.
\end{eqnarray}
The transformation rules can be found from the ${\cal N} = 8$ ones,
by making the above decomposition and by putting $\epsilon^8 = 0$.
In this way one finds
\begin{eqnarray}
\delta h_{\mu \nu} & = & \bar{\epsilon}^I \gamma_{(\mu} \psi^I_{\nu)} \,, \nonumber \\
\delta \psi^I_\mu & = & -\frac{1}{4} \gamma^{\rho \sigma} \partial_\rho h_{\mu \sigma} \epsilon^I - V_\mu^{IJ} \epsilon^J \,, \nonumber \\
\delta \psi_\mu & = & V_\mu^I \epsilon^I \,, \nonumber \\
\delta V_{\mu}^{IJ} & = &\frac{1}{2} \bar{\epsilon}^{[I} \phi^{J]}_\mu + \bar{\epsilon}^K \gamma_\mu \chi^{IJK} \,, \nonumber \\
\delta V_\mu^I & = & \frac{1}{4} {\bar\epsilon}^I\phi_\mu - \bar{\epsilon}^J \gamma_\mu \chi^{IJ} \,, \nonumber \\
\delta \chi^{IJK} & = & - \frac{3}{4} \gamma^\mu F_{\mu}^{[IJ} \epsilon^{K]} + \gamma^\mu \partial_\mu E^{IJKL} \epsilon^L + D^{IJKL} \epsilon^L \,, \nonumber \\
\delta E^{IJKL} & = & - \bar{\epsilon}^{[I} \chi^{JKL]} - \frac{1}{8} \epsilon^{IJKLMNO} \bar{\epsilon}^M \chi^{NO} \,, \nonumber \\
\delta D^{IJKL} & = & - \bar{\epsilon}^{[I} \dslash \chi^{JKL]} + \frac{1}{8} \epsilon^{IJKLMNO} \bar{\epsilon}^M \dslash \chi^{NO} \,, 
\end{eqnarray}
where $\phi_\mu = \phi^8_\mu$. The (anti-)self duality conditions
for $E^{IJKL}$ and $D^{IJKL}$ no longer hold but are instead
replaced by
\begin{eqnarray}
E^{IJKL} & = & \frac{1}{3!} \epsilon^{IJKLMNO} E^{MNO} \,, \nonumber \\
D^{IJKL} & = & -\frac{1}{3!} \epsilon^{IJKLMNO} D^{MNO} \,.
\end{eqnarray}
(One could as well remove $E^{IJK}$ and $D^{IJK}$ using these
rules.) The combination $E^{IJKL} D^{IJKL}$ is thus no longer zero
and can be added to the action. One obtains that the following
action is invariant
\begin{eqnarray}\label{linaction7}
S_3^{{\cal N}=7} & = & \int d^3x\,\bigg\{h^{\mu\nu}C_{\mu\nu}^{\rm
(lin)} + {\bar\psi}^I_\mu{\cal C}_{\rm (lin)}^{\mu\,I}
-2 V_\mu^{IJ}F_{\rm (lin)}^{\mu\,IJ} + \frac{8}{3} {\bar\chi}^{IJK}\chi^{IJK} \nonumber\\[.2truecm] & &- \frac{16}{3} E^{IJKL} D^{IJKL} - 8 \bar{\chi}^{IJ} \chi^{IJ} + 4 V_\mu^I F^{\mu I}_{\rm (lin)} - \bar{\psi}_\mu {\cal C}_{\rm (lin)}^{\mu} \bigg\}\,,
\end{eqnarray}
where the definitions of $F^{\mu I}_{\rm (lin)}$ and ${\cal C}_{\rm
(lin)}^{\mu}$ are the usual definitions of the (dual) field strength
of $V_\mu^I$ and Cottino tensor of $\psi_\mu$ respectively.

Adding the ${\cal N}=7$ supersymmetric LCS action to the ${\cal N}=8$ supersymmetric Einstein action
we obtain the action of ${\cal N}=7$ topologically massive supergravity:

\begin{eqnarray}
S^{{\cal N}=7}_{\text{TMG}} &=& \int\,d^3 x\,\bigg\{\frac{1}{2}
h^{\mu \nu} G^{(\mathrm{lin})}_{\mu \nu} +
\epsilon^{\mu\nu\rho}{\bar\psi}_\mu^i\partial_\nu\psi_\rho^i -
V_\mu^{ij}V_{\mu\,ij}
-\frac{2}{3} E^{ijkl}E_{ijkl} \nonumber\\[.2truecm]
&&  - \frac{4}{3} {\bar\psi}^{ijk}\partial \hskip -.24truecm
/\,\psi^{ijk}  +\frac{8}{3}   {\bar\psi}^{ijk}\chi_{ijk}
-\frac{4}{3} D^{ijkl}\Phi_{ijkl} +\frac{2}{3} \Phi^{ijkl}\Box \Phi_{ijkl}\bigg\}\nonumber\\[.2truecm]
&&+\frac{1}{\mu}\bigg\{h^{\mu\nu}C_{\mu\nu}^{\rm
(lin)} + {\bar\psi}^I_\mu{\cal C}_{\rm (lin)}^{\mu\,I}
-2 V_\mu^{IJ}F_{\rm (lin)}^{\mu\,IJ} + \frac{8}{3} {\bar\chi}^{IJK}\chi^{IJK} \nonumber\\[.2truecm]
& &- \frac{16}{3} E^{IJKL} D^{IJKL} - 8 \bar{\chi}^{IJ} \chi^{IJ} + 4 V_\mu^I F^{\mu I}_{\rm (lin)} - \bar{\psi}_\mu {\cal C}_{\rm (lin)}^{\mu} \bigg\}\,.
\end{eqnarray}
One may verify that the equations of motion lead precisely to a ${\cal N}=7$ spin 2 massive supermultiplet, see Table 1.
The analysis is similar to the ${\cal N}=4$ case treated in the  section \ref{sec:N=4}.


\section{Conclusions and Outlook}

The  correspondence between massless supermultiplets of 4D supersymmetry and massive supermultiplets of 3D supersymmetry, and the existence of
$\mathcal{N}$-extended 4D supergravity theories for $\mathcal{N}=1,2,3,4,5,6$ and $\mathcal{N}=8$ suggests the existence of  analogous parity-preserving 3D massive supergravity theories, with $\mathcal{N}$ now counting the number of 3D two-component Majorana spinor supercharges. In particular, the analogy suggests the existence of an $\mathcal{N}=8$ maximally supersymmetric extension of ``new massive gravity'' \cite{Bergshoeff:2009hq}. We also expect  additional parity-violating supergravity theories, such as ``topologically massive supergravity'' \cite{Deser:1982vy,Deser:1982sw} but  representation theory only allows
$\mathcal{N}\le7$ in this case.

The  general massive $\mathcal{N}=1$ 3D supergravity was constructed
in \cite{Andringa:2009yc,Bergshoeff:2010mf}, as was one version of
the linearized $\mathcal{N}=2$ 3D supergravity. A simplifying
feature of these unitary higher-derivative models is that the
higher-derivative terms have a quadratic approximation (in an
expansion about the Minkowski vacuum) that is invariant under
linearized superconformal gauge invariances. As a consequence, the
superconformal compensating multiplets needed for the construction
of generic higher-derivative  invariants are needed, {\it at the
linearized level}, only for the construction of the supersymmetric
extension of the Einstein-Hilbert term.  This means that we do not
need a full superconformal tensor calculus to construct the
linearized theories, which is fortunate because this has not been
worked out for $\mathcal{N}\ge3$ and must  involve an infinite
number of auxiliary fields for $\mathcal{N}\ge5$.

We have discussed the general picture of compensating fields for all $\mathcal{N}$. In particular $\mathcal{N}=3$ is similar to the $\mathcal{N}=2$ case discussed in \cite{Andringa:2009yc}, in that we only need one compensating multiplet which is a Stueckelberg multiplet in the sense that its coupling to the Weyl multiplet is bilinear. For $\mathcal{N} \ge 4$, we need multiplet copies of the compensating multiplet, and the Weyl multiplet couples to bilinears of them. The $R$-symmetry group gets spontaneously broken when we fix the scale transformations and the corresponding Goldstone bosons are the Stueckelberg fields.
We have presented the details of how things work out for $\mathcal{N}=4$, thereby constructing the general linearized massive 3D $\mathcal{N}=4$ supergravity, as well as a number of  (non-unitary) linearized models with yet higher derivative terms.

Maximally supersymmetric supergravity models are of particular interest, and we have constructed the linearized  maximally-supersymmetric  $\mathcal{N}=8$ ``new massive supergravity''.  The inclusion of the parity-violating Lorentz-Chern-Simons term necessarily breaks $\mathcal{N}=8$ supersymmetry, so that
$\mathcal{N}=7$ is maximal for models such as topologically massive gravity and we have constructed a linearized $\mathcal{N}=7$ topologically massive supergravity action.  However, it is not clear that  $\mathcal{N}=7$ is realizable beyond the linear level, because there are eight  Rarita-Schwinger fields in
the $\mathcal{N}=7$ multiplet; we think it likely that $\mathcal{N}=6$ is maximal for TMG.

The extension of our results to the full non-linear level is a challenging task that we leave to the future.  Given the striking properties of $\mathcal{N}=8$ supergravity in four dimensions, one may hope that the 3D  ``new massive''  $\mathcal{N}=8$ supergravity  will have similar nice properties. For example, a cosmological extension of the $\mathcal{N}=8$ massive gravity might allow an AdS vacuum preserving all 16 supersymmetries, in which case it might have a holographic dual 2D  conformal field theory with  maximal $(4,4)$ supersymmetry.


\section*{Acknowledgements}
We thank Paul Howe for helpful correspondence.
The work of OH is supported by the DFG -- The German Science Foundation
and in part by funds provided by the U.S. Department of Energy (DoE) under the cooperative research agreement DE-FG02-05ER41360. PKT thanks the
EPSRC for financial support.

\end{document}